\documentclass[twocolumn]{aastex631}
\usepackage{amsmath}
\usepackage{upgreek}
\usepackage{CJK}
\usepackage{graphicx}
\usepackage{subfigure}
\usepackage{enumerate}
\usepackage{multirow}
\usepackage{threeparttable}
\usepackage{booktabs}
\usepackage{hyperref}

\newcommand{\RNum}[1]{\uppercase\expandafter{\romannumeral #1\relax}}

\submitjournal{AAS Journals}

\shorttitle{Retrieval of planetary mass with N-body fitting model}
\shortauthors{Huang X.M. et al.}
\shortauthors{}

\graphicspath{{./}{figs/}}

\begin{document}
\begin{CJK*}{UTF8}{gbsn}

\title{Closeby Habitable Exoplanet Survey (CHES). \uppercase\expandafter{\romannumeral3}.  Retrieval of Planetary Masses in Binaries Using the N-body Model with RV and Astrometry Synergy.}

\correspondingauthor{Jianghui Ji}
\email{jijh@pmo.ac.cn}

\author{Xiumin Huang}
\affiliation{CAS Key Laboratory of Planetary Sciences, Purple Mountain Observatory, Chinese Academy of Sciences, Nanjing 210023, China}

\author{Jianghui Ji}
\affiliation{CAS Key Laboratory of Planetary Sciences, Purple Mountain Observatory, Chinese Academy of Sciences, Nanjing 210023, China}
\affiliation{School of Astronomy and Space Science, University of Science and Technology of China, Hefei 230026, China}
\affiliation{CAS Center for Excellence in Comparative Planetology, Hefei 230026, China}

\author{Chunhui Bao}
\affiliation{CAS Key Laboratory of Planetary Sciences, Purple Mountain Observatory, Chinese Academy of Sciences, Nanjing 210023, China}
\affiliation{School of Astronomy and Space Science, University of Science and Technology of China, Hefei 230026, China}

\author{Dongjie Tan}
\affiliation{CAS Key Laboratory of Planetary Sciences, Purple Mountain Observatory, Chinese Academy of Sciences, Nanjing 210023, China}
\affiliation{School of Astronomy and Space Science, University of Science and Technology of China, Hefei 230026, China}

\author{Su Wang }
\affiliation{CAS Key Laboratory of Planetary Sciences, Purple Mountain Observatory, Chinese Academy of Sciences, Nanjing 210023, China}
\affiliation{CAS Center for Excellence in Comparative Planetology, Hefei 230026, China}

\author{Yao Dong}
 \affiliation{CAS Key Laboratory of Planetary Sciences, Purple Mountain Observatory, Chinese Academy of Sciences, Nanjing 210023, China}
\affiliation{CAS Center for Excellence in Comparative Planetology, Hefei 230026, China}

\author{Guo Chen}
\affiliation{CAS Key Laboratory of Planetary Sciences, Purple Mountain Observatory, Chinese Academy of Sciences, Nanjing 210023, China}
\affiliation{School of Astronomy and Space Science, University of Science and Technology of China, Hefei 230026, China}
\affiliation{CAS Center for Excellence in Comparative Planetology, Hefei 230026, China}

\begin{abstract}

Given that secular perturbations in a binary system not only excite high orbital eccentricities but also alter the planetary orbital inclination,  the classical Keplerian orbital model is no longer applicable for orbital retrieval.  The combination of a dynamical model and observational data is essential for characterizing the configuration and planetary mass in close binaries. We calculate the theoretical radial velocity (RV) signal in the N-body framework and observe a drift in the RV semi-amplitude, which leads to a reduction in the $m$sin$i$ detection threshold by 20 $M_{\oplus}$,  with $\sim$ 100\% detection probability in the $m_1$sin$i_1$-$a_1$ parameter space.  High-precision RV data with an accuracy of 1 m/s can detect such dynamical effects. For four close-in binaries-GJ 86, GJ 3021, HD 196885, and HD 41004,  the deviation between the minimum mass derived from the Keplerian and N-body models is found to be $> 0.2 ~ M_{\mathrm{Jup}}$. High-precision astrometric data are also necessary to resolve the 3D orbits and true masses exoplanets. We generate astrometric simulation data with accuracies corresponding to Gaia (57.8 $\mu$as) and the Closeby Habitable Exoplanet Survey (CHES) (1 $\mu$as), respectively. Joint orbit fitting is performed for RV + Gaia and RV + CHES synergy methods. Compared with the fitting results from the astrometry-only model, the synergy models more effectively constrain the range of orbital inclinations. Using simulation data, we derive precise uncertainties for the true planetary mass, which are critical for determining the evolution of planets around binary stars and multi-planet systems.

\end{abstract}


\section{Introduction}\label{sec:1}

Advances in the precision and capability of exoplanet detection have enabled the identification of rocky planets around Sun-like stars. The latest generation of spectrographs, ESPRESSO (Echelle SPectrograph for Rocky Exoplanets and Stable Spectroscopic Observations) \citep{Pepe2010} and HARPS (High-Accuracy Radial velocity Planet Searcher) \citep{Wilken2010, Wilken2012, Cosentino2012}, achieve RV precision of up to 0.50 m/s. {While such extreme precision presents exciting opportunities, it also introduces challenges, such as the need to disambiguate between minuscule astrophysical signals and low-amplitude noise sources at the same scale.}

The orbital motions of interacting planets deviate from Keplerian dynamics. For close binaries containing planets and multi-planetary systems in MMRs, the classical Keplerian orbital model becomes inadequate for orbital fitting, as gravitational interactions drive high orbital eccentricities and alter planetary inclinations. \citet{Judkovsky2022} shows that when three planets are in a near-resonant chain, with the two super-periods close to one another, the Transit Timing Variations (TTVs) cannot be treated as system level TTVs due to resonance between the super-period signals. With higher precision of RV or photometric measurements, we are now able to detect signals arising from the underlying dynamics and provide more stringent constraints on planetary masses.

RV is not the only method capable of detecting gravitational interactions in planetary systems. \citet{Covarrubias2022} showed that perturbations resulting from N-body interactions can directly constrain planetary masses in multi-planet systems, while Keplerian orbits explain the majority of the astrometric motion of directly imaged planets. The synergy between astrometry and gravitational effects is exploited by the VLTI-GRAVITY instrument, which achieves astrometric precision up to 100 times better than existing methods \citep{Gravity_Collaboration2019}, a level of accuracy that enables detection of planet-planet interactions. In the HR 8799 system, \citet{Covarrubias2022} predicts planet-planet interaction deviations from Keplerian orbits \citep{Lacour2021} of up to 0.25 milliarcseconds within five years, making them detectable with VLTI-GRAVITY. The study suggests that using planet-planet interactions to measure dynamical masses may be more effective than relying on RV or absolute astrometry.

{In this work, we consider two primary types of non-restricted triple-body systems to evaluate the magnitude of dynamical effects: planets in mean-motion resonances (MMRs) and planets in close binaries. Approximately one-third of known multiple systems exhibit planets in low-order MMRs, and the dynamics of these systems provide valuable constraints on models of planetary migration. Currently, 154 binary star systems hosting planets have been detected, where a planet orbiting one of the two stellar components is classified as an S-type planet. If the separation between the binary components is relatively small ($<$100 au), the system is categorized as a close binary. In such systems, the mutual gravitational interaction between the stellar companion and the planet must be considered within the Jacobian framework, where the planet orbits the central star while the outer companion moves around the center of mass of the inner star-planet system. This approach more accurately represents the real dynamics of the system \citep{Lee2003}. Close binaries can significantly influence the formation and evolution of S-type planets through dynamical perturbations \citep{Xie2010,Gong2018}. Previous studies on the secular evolution of planets in close binaries have shown that extreme eccentricities and inclinations can be induced via the Eccentric Kozai-Lidov mechanism \citep{vonZeipel1910, Lidov1962, Kozai1962, Naoz2013, Li2014, Naoz2016} and secular chaos \citep{Rodet2021}.}

Dynamical astrometric fitting requires extremely high-precision astrometric data, which can be provided by the CHES mission \citep{Ji2022, Ji2024,Bao2024a,Bao2024b,Tan2024}. CHES is a space-borne astrometric mission designed to detect habitable planets around nearby solar-type stars (within $\sim 10$ pc) via micro-arcsecond relative astrometry. \citet{Bao2024a} (Paper I) considered the photocenter jitters induced by stellar activity in solar-type stars and found that the detection efficiency of planets in the habitable zones close to the stars is significantly reduced. In \citet{Tan2024} (Paper II), the effective observation strategy of CHES was detailed, outlining the relevant parameters for both target and reference stars.

After the extensive use of Keplerian orbital solution programs, such as \textit{EXOFIT} \citep{Balan2011} and \textit{orvara} \citep{Brandt2021}, several N-body orbit-fitting software packages have been recently developed and applied to RV and astrometric data analysis. \textit{PlanetPack} \citep{Baluev2013, Baluev2018} is a C++ software that includes user-friendly multi-Keplerian and Newtonian N-body RV fitting modules. \textit{Exo-striker} \citep{Trifonov2019}, constructed with a dynamical MCMC scheme, is designed to handle RV data and conduct long-term stability analysis of multi-planet systems. \citet{Judkovsky2022} developed the analytical code \textit{AnalyticLC} to model the dynamics of planetary systems and calculate light-curve, RV, and astrometric signatures in 3D. \citet{Covarrubias2022} suggested that using planet-planet interactions to measure dynamical masses might be a more effective method than relying on radial velocities and/or absolute astrometry. They also incorporated the N-body integration module REBOUND \citep{Rein2012, Rein2015} into the Python orbit-fitting program \textit{orbitize!}. In this work, we present a comprehensive Python package that not only performs theoretical analysis of orbital RV and astrometry signals derived from the N-body model but also provides dynamical orbital solutions for planets in close binaries, leveraging both RV and astrometry measurements.

To further demonstrate that the effects of gravitational perturbations can be detected, we calculated the theoretical RV detectability within the N-body framework. From a theoretical standpoint, gravitational perturbations from the outer companion induce the variation of the planetary eccentricity, leading to a drift in the RV signal of the host star. This drift can be detected with an RV accuracy of 1 m/s. The drift in RV semi-amplitude $K$ significantly impacts the detection efficiency of planetary signals in both MMRs and binary systems, highlighting the importance of integrating N-body dynamics with observations.

To better constrain the measurement uncertainty of planetary masses, we develop the N-body orbital retrieval program as a python package capable of solving orbits in six cases: two-Keplerian RV model, N-body RV model, two-Keplerian astrometry model, N-body astrometry model, two-Keplerian RV+Astrometry synergy model, and N-body RV+Astrometry synergy model. The diversity of fitting models expands the range of possible values for planetary orbital elements and masses, providing information for the classification of planets. We demonstrate that the N-body solutions for the minimum planetary mass are smaller than those derived from the Keplerian model by $> 0.2$ $M_{\mathrm{Jup}}$ for the systems GJ 86, GJ 3021, HD 196885, and HD 41004. In comparison with the RV+Astrometry synergy model, and accounting for system stability, the CHES astrometry fitting results yield the best goodness of fit for GJ 86 Ab.

In Section \ref{sec:2}, theoretical analysis predicts that the RV semi-amplitude $K$  undergoes a drift due to the presence of an outer companion, which in turn impacts the detection efficiency. Section \ref{sec:3} focuses on the N-body RV fitting of S-type planets in close-in binaries, where dynamical fitting results yield smaller minimum planetary masses compared to the Keplerian model. In Section \ref{sec:4}, the synergy between RV and high-precision astrometry simulation data is presented, and the true planetary mass of GJ 86 Ab is derived, showing a difference of $0.5 \sim 0.6 ~ M_{\mathrm{Jup}}$ between the Astrometry-only method and the RV + Astrometry synergy model. In Section \ref{sec:5}, we summarize the major conclusions from each section, including the variation in detection probability, the best-fitting results of the RV N-body model, and the RV+Astrometry synergy method.

\section{Detectability of RV with N-body Model }\label{sec:2}

From an observational perspective, the semi-amplitude of the RV can be derived from the observed stellar spectral data and the Doppler shift of the frequency. Additionally, the relative motion velocity between the planet and the star can be theoretically calculated using Keplerian orbital dynamics. This Section compares the theoretical RV signal with the detection threshold of spectrographs over observation time to assess the detection efficiency of RV within the N-body model. The semi-amplitude of the RV can be calculated as follows:

\begin{equation}
 \begin{aligned}
\label{equ:RV_K}
&K_{1}=\left(\frac{2 \pi G}{P_{1}}\right)^{1 / 3} \frac{m_{1} \mathrm{sin} i_{1}}{\left(m_{0}+m_{1}\right)^{2 / 3}} \frac{1}{\sqrt{1-e_{1}^{2}}}
 \end{aligned}
\end{equation}
where $P_{1}$, $i_{1}$ and $e_{1}$ are Keplerian orbital period,  orbital inclination, and eccentricity.  If we assume the orbital inclination is $90^{\circ}$,  and the planetary mass is much smaller than the primary star,  the simplified minimum mass is approximated to be:
\begin{equation}
 \begin{aligned}
\label{equ:min_m}
m_{1}=K_{1}\left(\frac{P_{1}}{2 \pi G}\right)^{1 / 3}{m_{0}^{2 / 3}}\sqrt{1-e_{1}^{2}}
 \end{aligned}
\end{equation}

In the Keplerian orbital model, the parameters $P_{1}$, $i_{1}$, and $e_{1}$ remain constant over time. The detectability of $K_1$ is therefore constrained by the resolution of the spectrometer. In more common scenarios, where massive planets or stellar companions are present in the system, the Keplerian orbit evolves. The time-series variation of $K_1$ shows a tiny drift. When the magnitude of the tiny drift becomes comparable to the accuracy of high-precision spectrometer,  RV data can reveal detailed dynamical signals from planetary systems. Next, we demonstrate that high-precision RV data, simulated with the precision of ESPRESSO ($\sim 1$ m/s), can detect dynamical perturbations in binary and mean-motion resonant systems.

\subsection{Mean Motion Resonances}\label{subsec:2.1}

\begin{figure*}
\begin{center}
       \subfigure[]{\includegraphics[width=0.85\columnwidth,height=6.5cm]{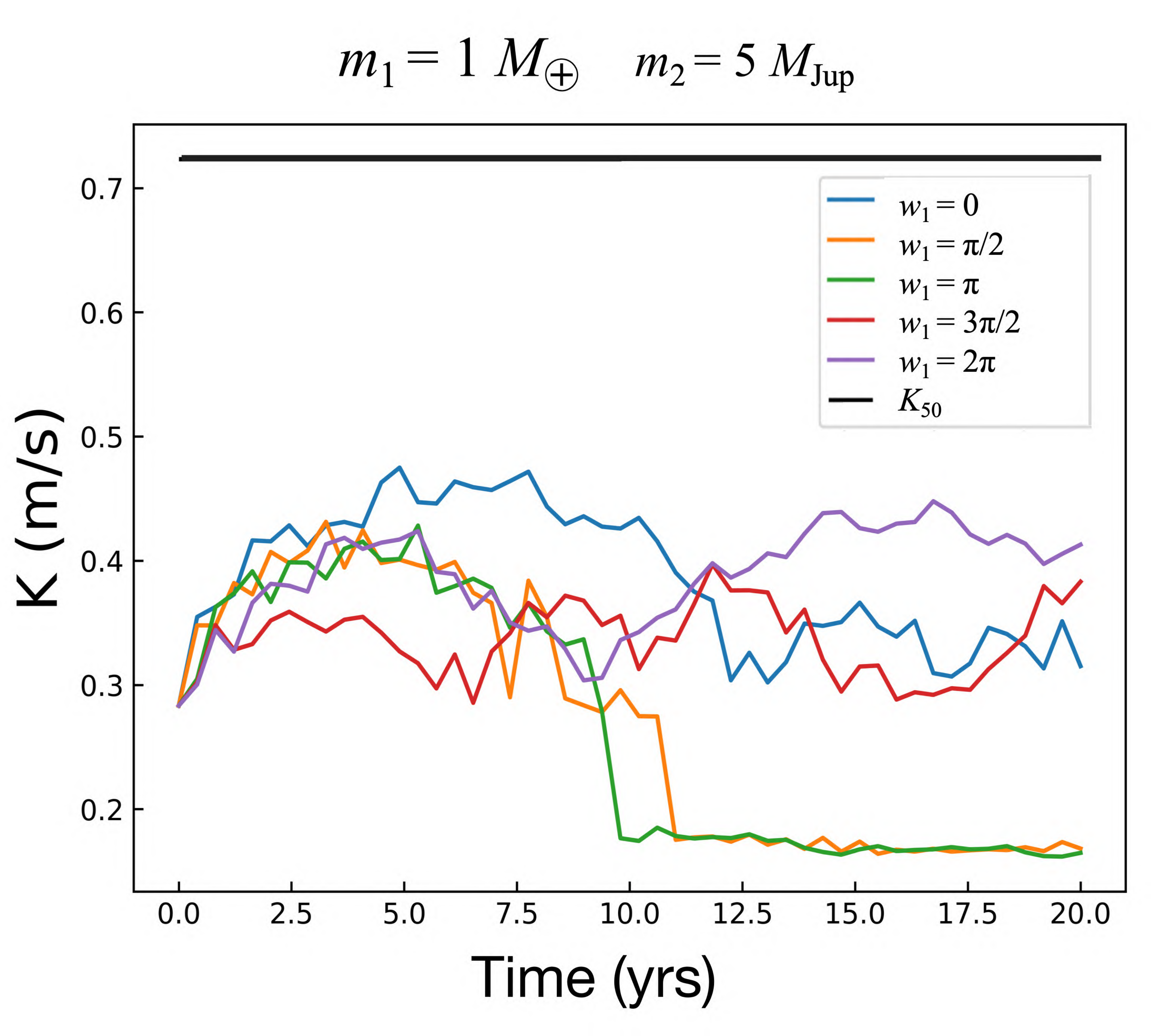}}
         \subfigure[]{\includegraphics[width=0.85\columnwidth,height=6.5cm]{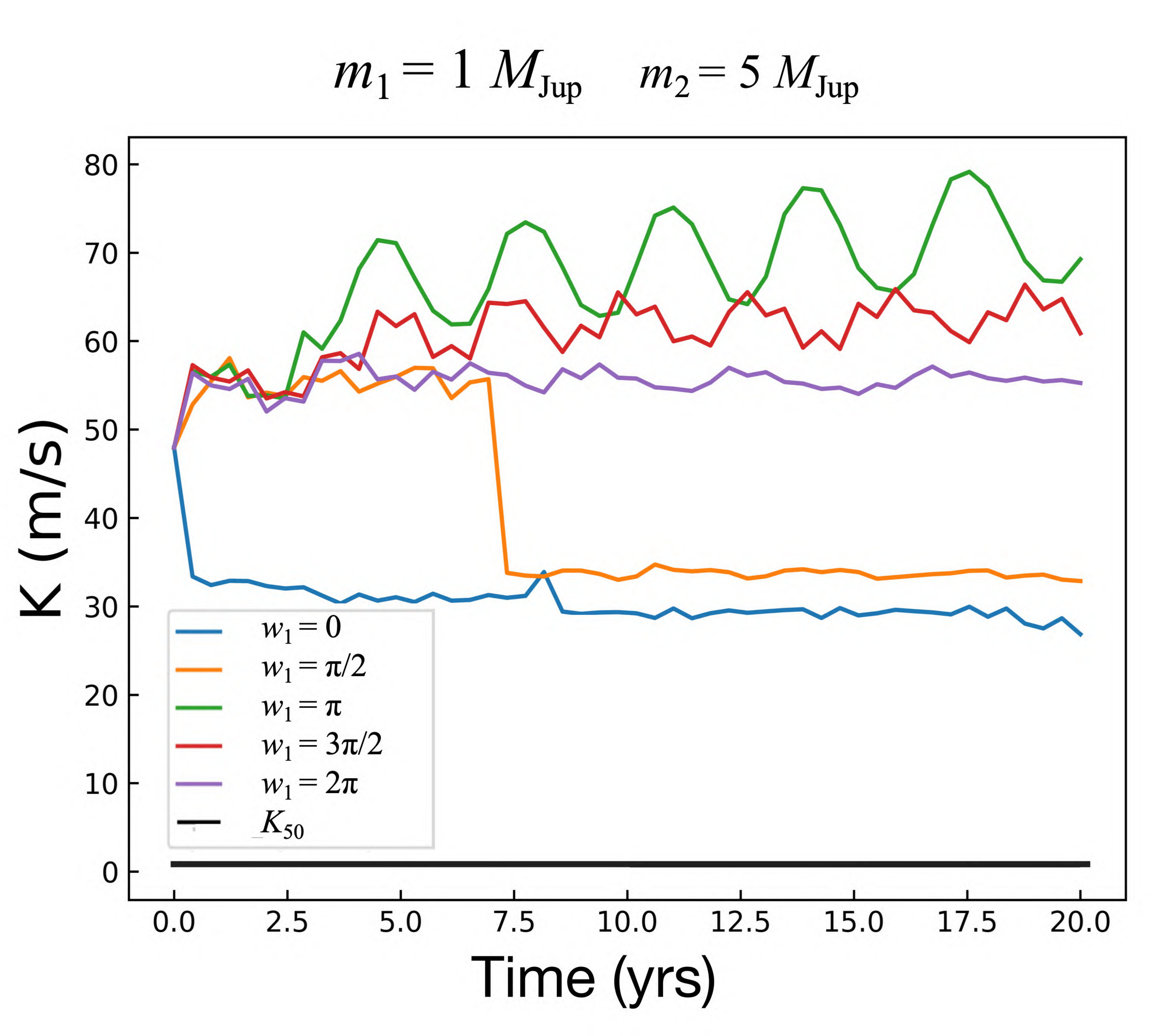}}
    \caption{Evolution of the RV semi-amplitude $K$ with time,  and the drift of $K$ for the Earth-Jupiter (left) and Jupiter-Jupiter (right) systems in the 2:1 MMRs orbital configuration during the RV observations baseline of 20 years. The solid black line is the RV threshold corresponding to the 50\%  detection efficiency. Colored lines represent cases with different initial arguments of periastron \( \omega_1 = 0 \text{-} 2\pi \). Both panels show a clear dispersion of \( K \) for different initial \( \omega_1 \). The total detection probability increases when \( K \) drifts upward and exceeds the \( K_{\mathrm{50}} \) threshold in Panel (a); otherwise, it decreases when \( K \) drifts downward below \( K_{\mathrm{50}} \) in Panel (b).
}
    \label{fig:rv_K_mmn}
    \end{center}
\end{figure*}

\begin{figure*}
\begin{center}
       \subfigure[]{\includegraphics[width=1.0\columnwidth,height=7cm]{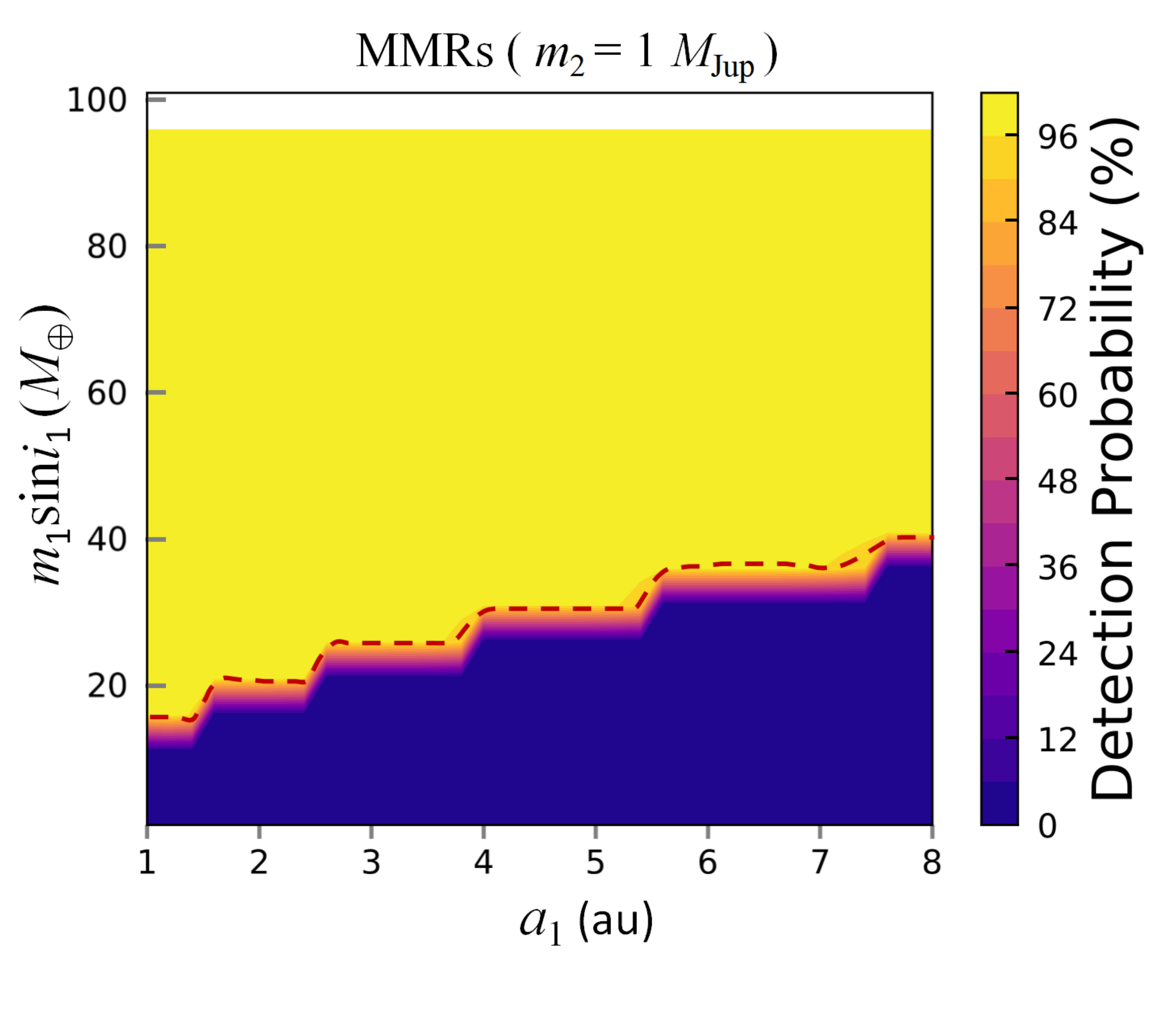}}
          \subfigure[]{\includegraphics[width=1.0\columnwidth,height=7cm]{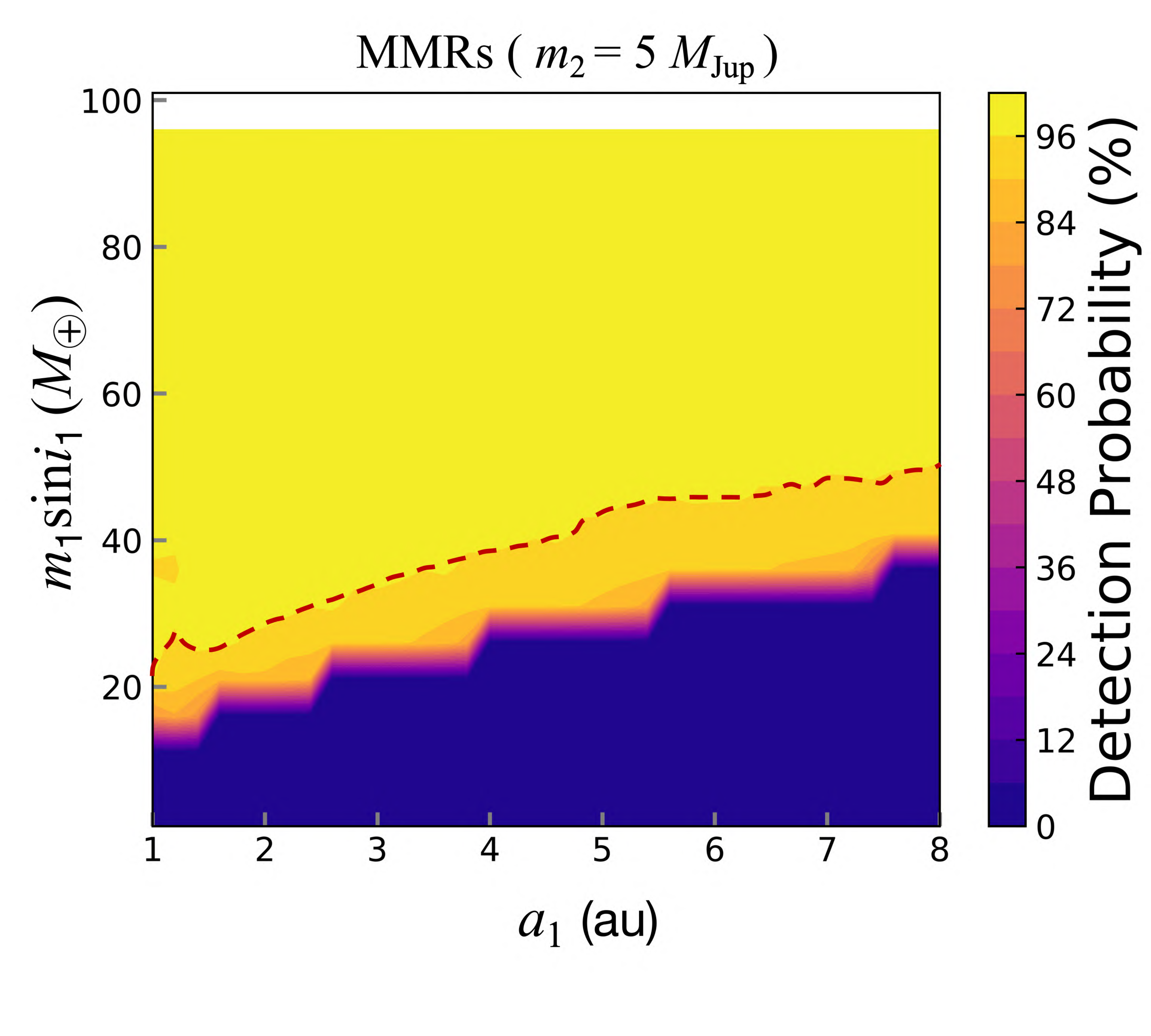}}\\
          \subfigure[]{\includegraphics[width=1.0\columnwidth,height=7cm]{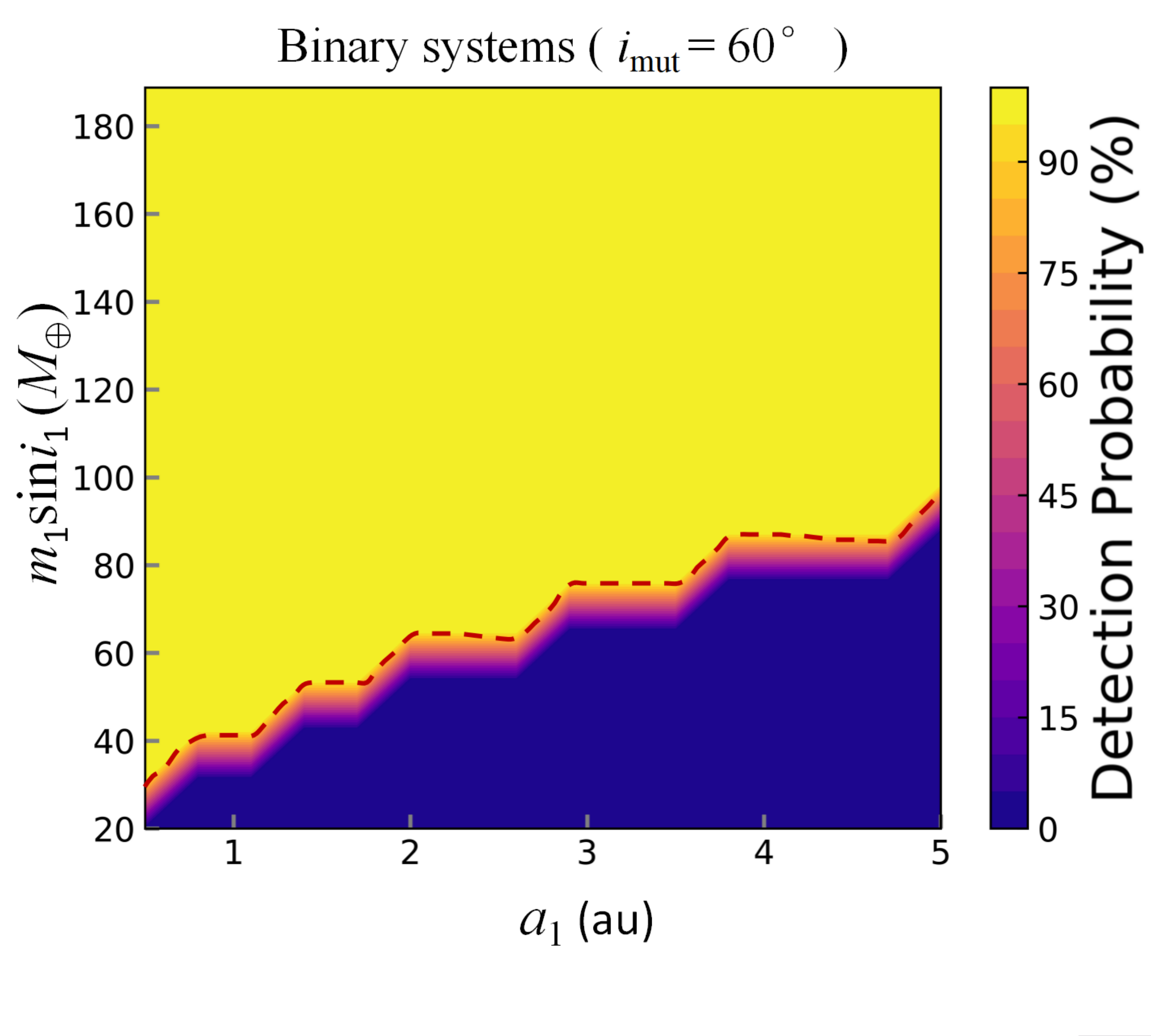}}
         \subfigure[]{\includegraphics[width=1.0\columnwidth,height=7cm]{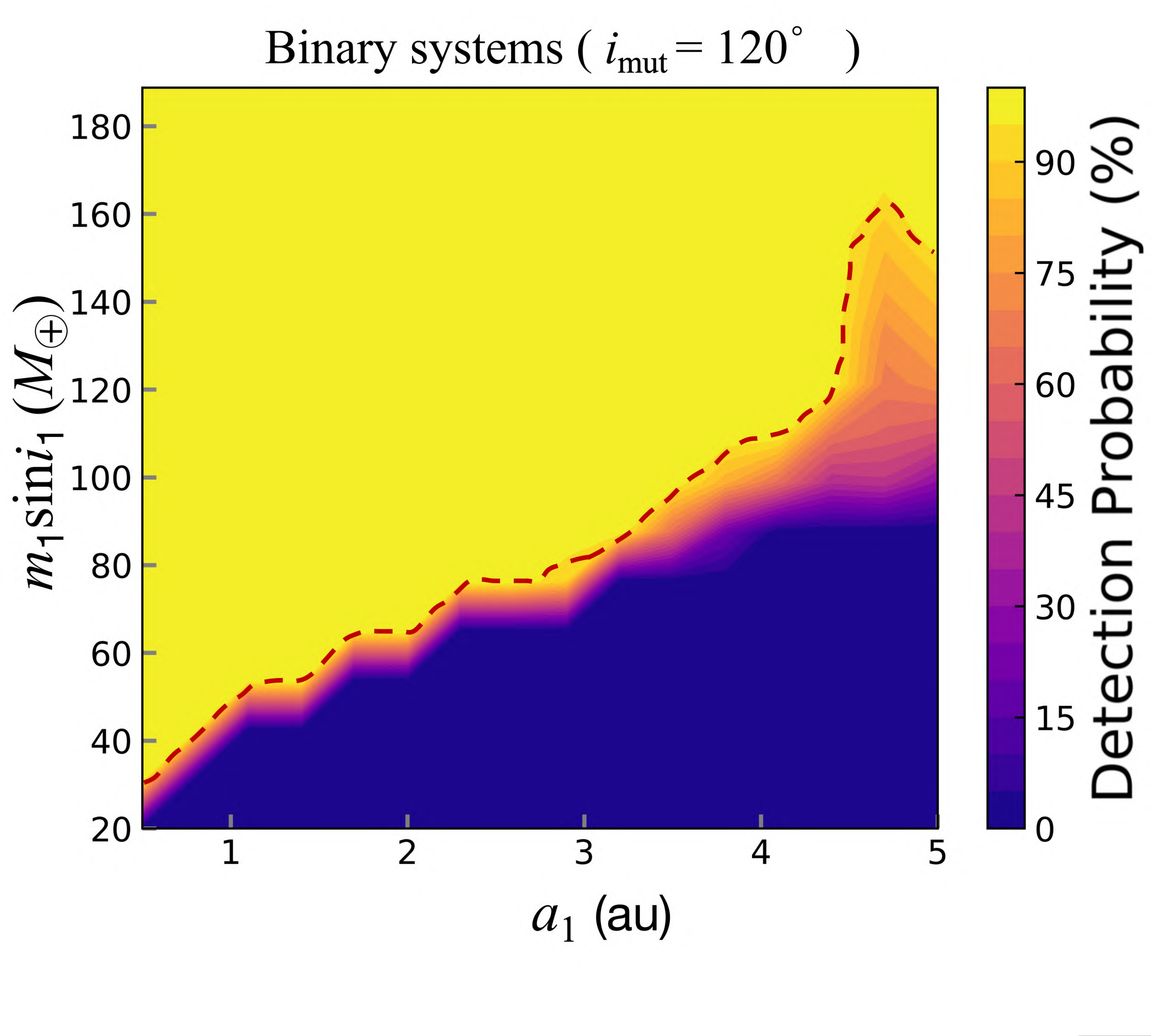}}
    \caption{Detection probability in the parameter spaces of $m_1\sin i_1$ -- $a_1$ for MMRs systems and close-binaries. The upper two panels show the influence of the MMRs on the detectability of the smallest planetary masses with an RV precision of 1 m/s and 100 observations.  The perturbation object in Panel (a) has a mass of 1 Jupiter mass, whereas in Panel (b), it has a mass of 5 Jupiter masses. The below two panels show the detectability of planets in binary systems based on the N-body RV model.  The mutual inclination in Panel (c) is $60^{\circ}$, which in Panel (d) is $120^{\circ}$. {The red dashed line marks the boundary between regions of $>$ 95\% probability and those with lower probability. This boundary shifts upward in Panels (b) and (d) as compared to the corresponding plots in the left column.} Panel (b) indicates a uniform decrease in the total detection probability between $a_1$ = 1 -- 8 au, while in Panel (d), the decrease of total detection probability is indicated by a local bump between $a_1$ = 4 -- 5 au.}
    \label{fig:rv_efficiency_binary}
    \end{center}
\end{figure*}

Jupiter's Galileo satellites are known to be in Laplace resonance, where the ratio of their orbital periods defines the resonance order. By analyzing data from the Kepler telescope over several years, researchers have discovered that a significant number of multi-planetary systems exhibit mean-motion resonant configurations. Among the Kepler multi-planet systems, the 2:1 and 3:2 MMRs are observed to be more common. The distribution of period ratios between adjacent Earth-like planets observed by \textit{Kepler} Mission reveals that planetary MMRs tend to cluster around the ratios 3:2, 2:1, 5:2, and 3:1.

According to \citet{Narayan2005},  the RV semi-amplitude $K$ from a closely orbiting several Earth-mass object is given by:
\begin{equation}
\label{equ:KEarth}
K=6.4 \mathrm{~m} \mathrm{~s}^{-1}\left(\frac{m_\mathrm{p}\sin i}{10 ~M_{\oplus}}\right)\left(\frac{P}{\text { days }}\right)^{-1 / 3}\left(\frac{M_*}{M_{\odot}}\right)^{-2 / 3}
\end{equation}

From the perspective of forward derivation, the theoretical RV signals are calculated across the parameter spaces of planetary mass $m_\mathrm{p}$, orbital period $P$, RV measurement precision $\sigma$, and the number of observations $N$. Following \citet{Cumming2004}, the detection threshold for $K$ is derived from the signal-to-noise ratio required to detect the signal:

\begin{equation}
\label{equ:signal_noise}
\begin{aligned}
\frac{K}{\sqrt{2} \sigma} & =\left[\left(\frac{N_i}{F}\right)^{2 /(N-3)}-1\right]^{1 / 2} \\
& \approx\left[\frac{2 \ln \left(N_i / F\right)}{N}\right]^{1 / 2},
\end{aligned}
\end{equation}
where $N_i$ is the number of independent frequencies searched, $F$ denotes the specified false alarm probability. Thus the critical condition of $K$ that can be detectable in 50$\%$ of observation time is derived as:

\begin{equation}
\label{equ:K50}
K_{50}=\frac{6 \mathrm{~m} \mathrm{~s}^{-1}}{\sqrt{N}}\left(\frac{\sigma}{\mathrm{~m} \mathrm{~s}^{-1}}\right)\left[\frac{\ln \left(N_i / F\right)}{9.2}\right]^{1 / 2}
\end{equation}

The planetary mass with a detectable efficiency of $50\%$ is:
\begin{equation}
\label{equ:M50}
\begin{aligned}
M_{50} &\approx \frac{10 M_{\oplus}}{\sqrt{N}}\left(\frac{\sigma}{\mathrm{~m} \mathrm{~s}^{-1}}\right)\left(\frac{P}{\text { days }}\right)^{1 / 3}\left[\frac{\ln \left(N_i / F\right)}{9.2}\right]^{1 / 2} \\
&*\left(\frac{M_*}{M_{\odot}}\right)^{2 / 3}
\end{aligned}
\end{equation}

First, we apply the N-body RV model to multi-planetary systems that include a Jupiter-like planet and a hot Earth located near the 2:1 MMRs. The simulation setup is as follows: the semi-major axis of the inner terrestrial planet's orbit ranges from 0.1 au to 0.8 au, with the mass of the adjacent Jupiter set between 1 and 7 $M_{\mathrm{Jup}}$. The two planetary orbits are assumed to be coplanar, with $e_1 = 0$ and $e_2 = 0.2$. We utilize the REBOUND  \citep{Rein2012} N-body module in our python package and compute the inner perturbed orbit with IAS15 integrator \citep{Rein2015} . The semi-amplitude of the RV, $K$, is calculated using the N-body model. Since the perturbed orbit evolves over time, $K$ varies with the observation time. The evolution curves of $K$ for $a_1 = 0.1$, 0.172, 0.244, 0.316, and 0.388 au show dispersion with different arguments of periastron $\omega$, with the dispersion of $K$ increasing with the mass of the Jupiter-like planet.

The planetary mass and orbital position of the inner orbit are selected such that the theoretical apparent velocity signal is comparable to the magnitude of $K_{\mathrm{50}}$ with a detection accuracy of 1 m/s, which motivates the choice of a hot Earth. The outer planet resides in a 2:1 MMRs with the hot Earth, such that $a_2$ corresponds to $a_1$. Regarding the choice of eccentricity, during the process of orbital migration, the MMRs is typically excited by the outer planet, and the eccentricity $e_2$ plays a significant role in the resonance's influence on the orbital dynamics, while $e_1$ is typically near-circular prior to the excitation of the MMRs.

We further compare the detection efficiency $K_{\mathrm{50}}$ with $K$. Considering the ESPRESSO's accuracy, the RV precision in Equation \ref{equ:K50} is set to $\sigma = 1$ m/s, and the number of observation is set to $N = 100$ over 20 years. Additionally, we set $N_i = 8$ and $F = 0.01$. Figure \ref{fig:rv_K_mmn} shows the evolution of the RV semi-amplitude $K$ with time, in Figure \ref{fig:rv_K_mmn} (a), the simulated values of $K$ are consistently below $K_{\mathrm{50}}$ for Earth-mass planets, whereas the maximum $K$ will approach the 50\% detection threshold as the mass of the Jupiter-like planet increases.

Next, we replace the inner perturbed planet with a Jupiter-mass planet, keeping the other simulation parameters unchanged. We plot the evolved $K$ and the detection threshold $K_{\mathrm{50}}$ simultaneously, showing that $K$ eventually crosses below the $K_{\mathrm{50}}$ line. The results for the inner planet are presented in Figure \ref{fig:rv_K_mmn} (b).

Secondly, if the minimum value of $K_{\mathrm{min}}$ during the 20-year observation time remains greater than 1 m/s, we consider the detection efficiency to be 100\%. For the simulation cases, we calculate the initial RV semi-amplitude $K_0$ at the first epoch of the 20-year observation baseline. It is always found that $K_0 > 1$ m/s for super-Earths and Neptune-like planets. For systems where $K_0 > 1$ m/s, upward-drifting cases do not affect the overall detection probability, while downward-drifting cases lead to a decrease in detection probability.
For sub-Earth with $K_0 < 1$ m/s,  upward-drifting cases could significantly increase the detection efficiency and detection probability.
Figure \ref{fig:rv_efficiency_binary} shows the detection probability map with 100\% efficiency for Earth-like and Jupiter-like planets in the $m_1$-$a_1$ parameter space.

Simulation setup for Figure \ref{fig:rv_efficiency_binary} (a) and (b): $m_1 \sin i_1 = 1 - 100~M_{\oplus}$, $m_2 = 1~M_{\mathrm{Jup}}$ in (a), $m_2 = 5 ~M_{\mathrm{Jup}}$ in (b), $a_1 = 1 - 8.2$ au, with $a_2$ located at the 2:1 MMRs relative to $a_1$. The  eccentricities are set as $e_1 = 0.05$ and $e_2 = 0.35$, with $\omega_1 = 0 - 2\pi$ and $\omega_2 = 0$. Both orbits are assumed to be coplanar.

The yellow region in Figure \ref{fig:rv_efficiency_binary} represents detection probabilities greater than 95\%, with the lower boundary corresponding to the 95\% detection threshold of $m \sin i$. A comparison between panels (a) and (b) exhibits that the total detection probability is reduced due to RV drift. As the perturbation is enhanced, more cases drift downward, falling below the RV detection criterion of 1 m/s. This results in a decrease of the $m \sin i$ detection threshold by 20 $M_{\oplus}$ in the $m_1$sin$i_1$-$a_1$ parameter space. Further simulations also show that higher-order MMRs lead to more significant $K$ drifts.

\subsection{Close Binary Systems}\label{subsec:2.2}
We selected the $\gamma$ Cep system as a case study to quantify the gravitational effects of a stellar companion on the theoretical RV values in binary systems. The RV signal of the planet in $\gamma$ Cep was first measured by \citet{Campbell1988}. This close binary system, located at a distance of 13.79 pc \citep{Hatzes2003}, is also a candidate target for the future high-precision astrometric mission CHES \citep{Ji2022, Ji2024}. The primary star, $\gamma$ Cep A, is a K1III-IV star with a stellar mass of $1.40 \pm 0.12~ M_{\odot}$ \citep{Neuhauser2007}. \citet{Neuhauser2007} directly detected the companion star $\gamma$ Cep B, whose orbital elements are $m_2 = 0.409 \pm 0.018~M_{\odot}$, $a_2 = 20.18 \pm 0.66$ au, $i_2 = 119.3^{\circ}$, and $\Omega_2 = 18.04^{\circ} \pm 0.98$. The planet $\gamma$ Cep Ab orbits nearly perpendicularly to the binary system \citep{Reffert2011}.

\citet{Huang2022} reported two set of solutions derived by N-body MCMC fitting in the Jacobian reference frame for $\gamma$ Cep Ab: the RV semi-amplitude $K_1 = 28.08^{+1.23}_{-1.45}$ m/s,   $a_1 =2.1459\pm0.0048$ au,  $e_1 = 0.0724^{+0.0575}_{-0.0879}$,  $\omega _1= {48.47^{\circ}}^{+1.81}_{-4.50}$,  and the epoch of periastron  passage $T_{\mathrm{p,1}} = JD-2453140.16 ^{+38}_{-34}$.  With the planetary inclination $i_1 = 5.7^{\circ}$, as reported by \citet{Reffert2011}, the planetary mass is estimated to be $17.58 \pm 0.7$ $M_{\mathrm{Jup}}$.

We adopted the orbital elements of $\gamma$ Cep Ab as reported above and transformed them into Cartesian coordinates and velocities for the first observational epoch. Using these initial conditions, we employed the N-body integrator IAS15 \citep{Rein2015} and the Newtonian motion equations to calculate the orbits of both the planet and the secondary star over the course of the observation period. A comparison of the RV signals reveals that the deviation in the stellar radial velocities between the Keplerian and N-body models approaches 20 m/s, yielding a relative error of approximately $10^{-2}$. The deviation in the radial velocities induced by the planet between the Keplerian and N-body models reaches 5 m/s, corresponding to a relative error of about $10^{-1}$.

We apply the N-body RV model to more general close-binary systems, assuming the RV spectrometer precision to be 1 m/s, consistent with the precision of the current high-precision spectrograph ESPRESSO. The simulation setup for Figure \ref{fig:rv_efficiency_binary} (c) and (d) is as follows: the stellar mass $M_{\star} = 1.4~M_{\odot}$, $m_1 \sin i_1 = 20 - 200~M_{\mathrm{Jup}}$, $m_2 = 0.4~M_{\odot}$, $a_1 = 0.5 - 5.3$ au, $a_2 = 18$ au, $e_1 = 0.05$, $e_2 = 0.35$, with $\omega_1 = 0 - 2\pi$, and $\omega_2 = 0$. We treat the mutual inclination $i_{\mathrm{mut}}$ of $m_1$ and $m_2$ as a free parameter, with $i_{\mathrm{mut}}$ ranging from 0$^{\circ}$ to 180$^{\circ}$. In Figure \ref{fig:rv_efficiency_binary} (c), we set $i_{\mathrm{mut}} = 60^{\circ}$, and in Figure \ref{fig:rv_efficiency_binary} (d), $i_{\mathrm{mut}} = 120^{\circ}$. We present a diagram showing the detectability of the minimum planetary mass as a function of semi-major axis and mutual inclination. Clearly, between $a_1 = 4$ au and $a_1 = 5$ au, there is a local decrease in detection probability, with the $m \sin i$ detection threshold increasing by approximately 80 $M_{\oplus}$. This is primarily driven by the drift of $K$ in close-binary systems. Thus, variations in the observed RV signal can provide insights into the configuration and dynamical stability of these systems.

\begin{table*}
\centering
\caption{Stellar parameters of the primary star in six close-binaries}
\label{tab:close_binary}
\begin{threeparttable}[b]
\begin{tabular}{lcccccc}
\toprule
Parameter & GJ 86$^{1}$ & $\tau$ Bootis$^{2}$& GJ 3021$^{3}$& HD 196885$^{4}$& HD 41004$^{5}$& HD 164509$^{6}$\\
\midrule
Spectral type & K1V &  F6V & G6V & F8V &KIV & G5V\\
Distance (pc)& 10.9 &  $15.6521\pm0.0843$& $17.62\pm0.17$& $33.0\pm1.0$ &$43.03\pm{1.81}$ &$52.0\pm3.0$\\
$T_{\mathrm{eff}}$ (K) &  5350 & $6466.27\pm106.03$ & $5540.0\pm75.0$& $6340.0\pm39.0$ & 5010 &  $5922.0\pm44.0$\\
$[Fe/H]$ & -0.24 &  $0.264\pm0.020$& $0.10\pm0.08$& $0.29\pm 0.05$ &$-0.09\pm{0.10}$&  $0.21\pm0.03$\\
$M_{\star}(M_{\odot})$&  $0.88\pm0.12$& $1.32\pm0.21$& 0.9 & 1.33 & 0.7&$1.13\pm0.02$\\
$R_{\star} (R_{\odot} )$&  $0.7905\pm0.0519$& $1.426\pm0.057$& 0.9 & $1.79\pm0.17$& --&$1.06\pm0.03$\\
Age (Gyr)& 2.03 &  $1.3^{+0.4}_{-0.6}$& 8.77 & $2.0\pm0.5$ & 1.64&$ 1.1\pm1.0$\\
\bottomrule
\end{tabular}
 \begin{tablenotes}[para,flushleft]
        \footnotesize
        \item References: 1, \citet{Flynn1997,Queloz2000,Els2001}; 2, \citet{Stassun2019}; 3,\citet{Rocha-Pinto1998,Naef2001,Chauvin2006}; 4, \citet{Correia2008},  5, \citet{Santos2002,Zucker2004}, 6, \citet{Wittrock2016}.
  \end{tablenotes}
  \end{threeparttable}
\end{table*}

\begin{figure*}
\begin{center}
        \subfigure[]{\includegraphics[width=0.9\columnwidth,height=7cm]{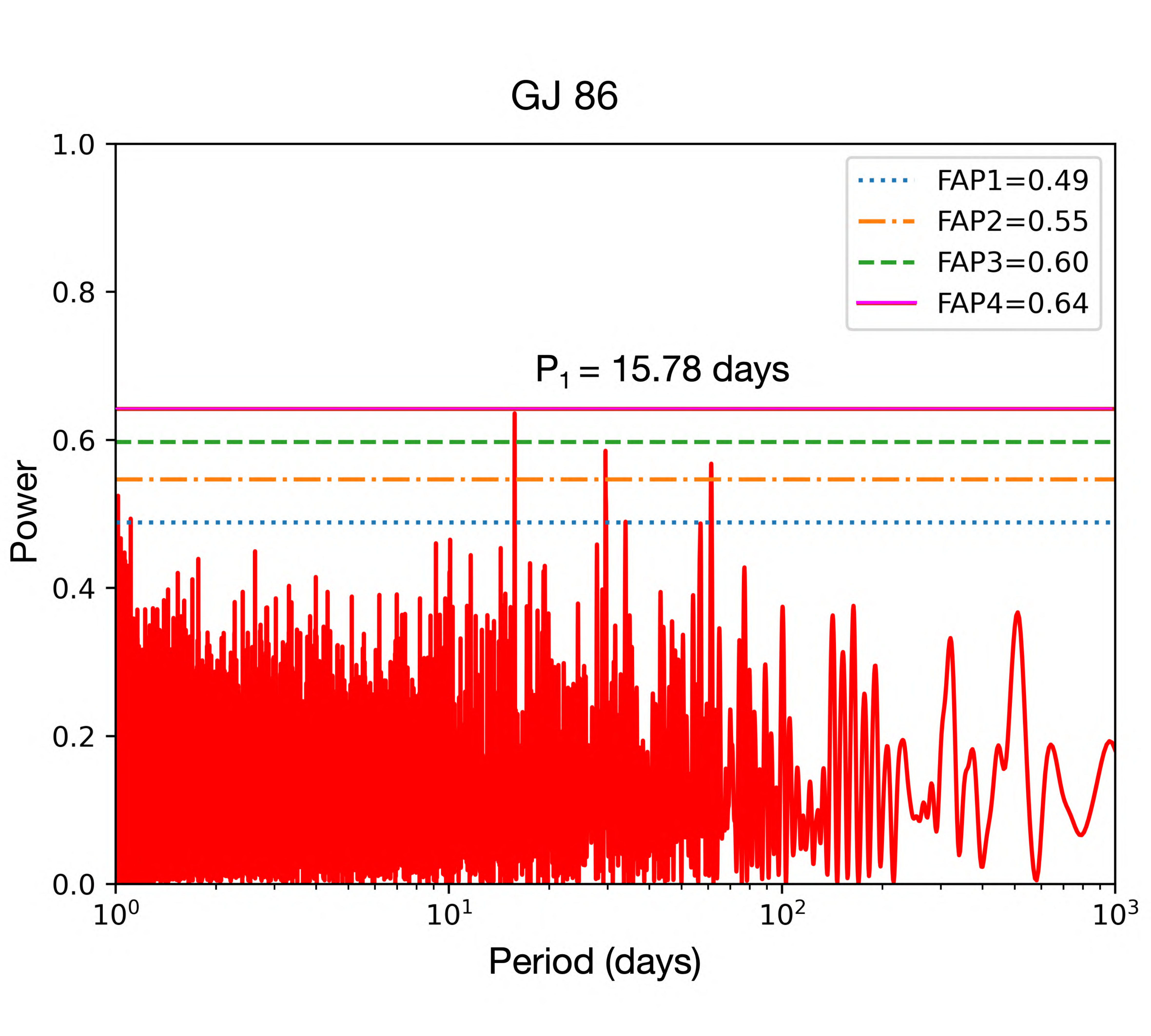}}
         \subfigure[]{\includegraphics[width=0.9\columnwidth,height=7cm]{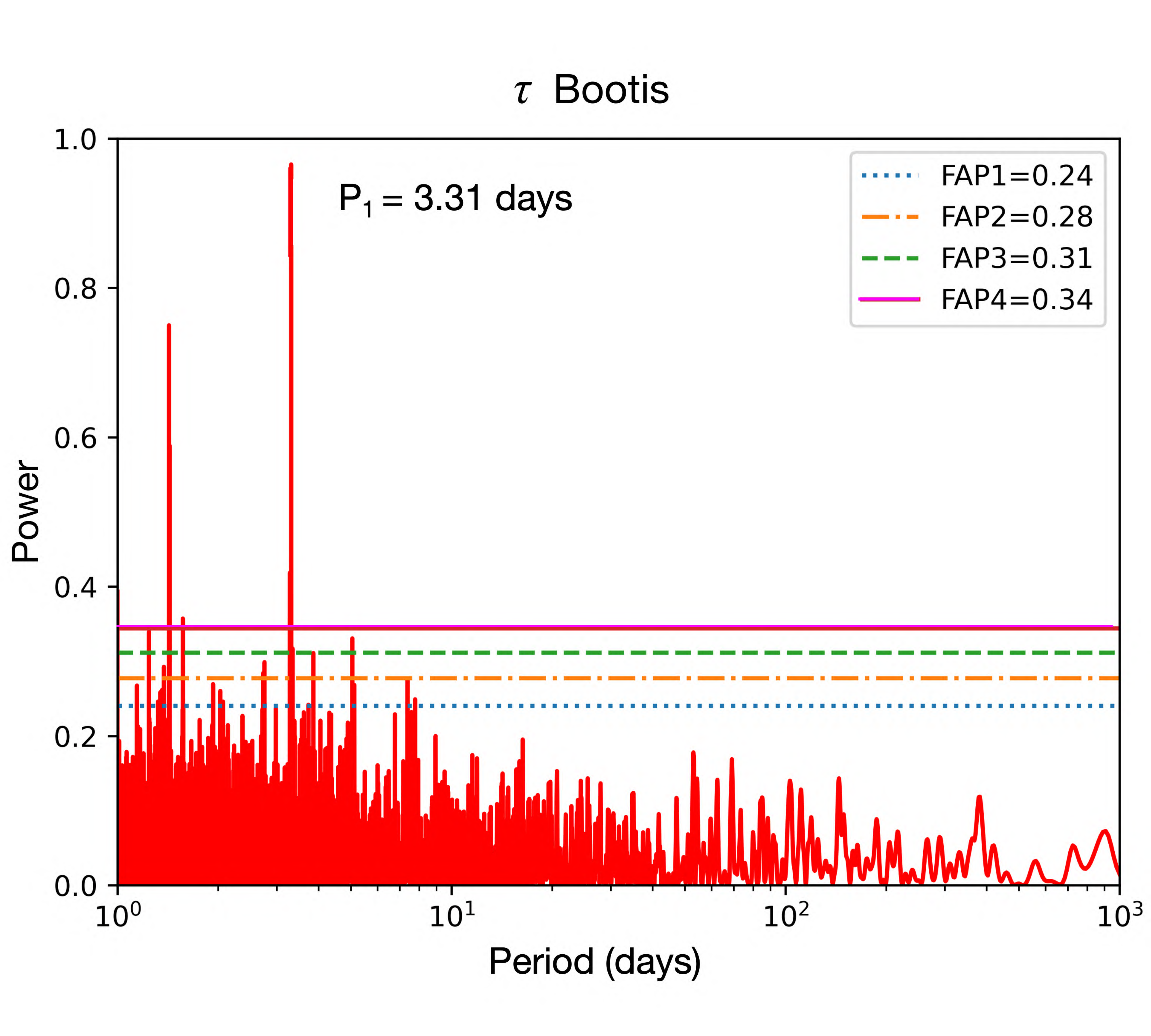}}
  \caption{The Lomb-Scargle periodogram of GJ 86 Ab and $\tau$ Bootis Ab. The value of the FAP represents the maximum power of the periodic signals. FAP1, FAP2, and FAP3 in the legend correspond to false alarm probabilities of 1\%, 0.1\%, and 0.01\%, respectively. Significant periodic signals are observed at \( P_1 = 15.78 \) days for GJ 86 Ab and \( P_1 = 3.31 \) days for \( \tau \) Bootis Ab, indicating that these signals are truly periodic.
}
  \label{fig:ls}
  \end{center}
\end{figure*}

\section{RV N-body Fitting of Close Binary Systems}\label{sec:3}

With the advancement of ground-based high-precision RV spectrometers and space-based detection capabilities, the focus of exoplanet science is gradually shifting from large sky surveys to the characterization of planetary systems, planetary physical parameters, and planetary habitability. Planets orbiting nearby stars are prime targets for astrometric measurements of their properties. These nearby stars are also potential candidates for future missions such as the Habitable Worlds Observatory (HWO) \citep{Dressing2024}, which aims to directly image terrestrial planets. In this study, we compiled a list of close binary systems and selected six systems for N-body fitting of RV data: GJ 86, $\tau$ Bootis, GJ 3021, HD 196885, HD 41004, and HD 164509, ranked by their distance from the Solar System.

\begin{figure*}
\begin{center}
        \subfigure[]{\includegraphics[width=0.9\columnwidth,height=7cm]{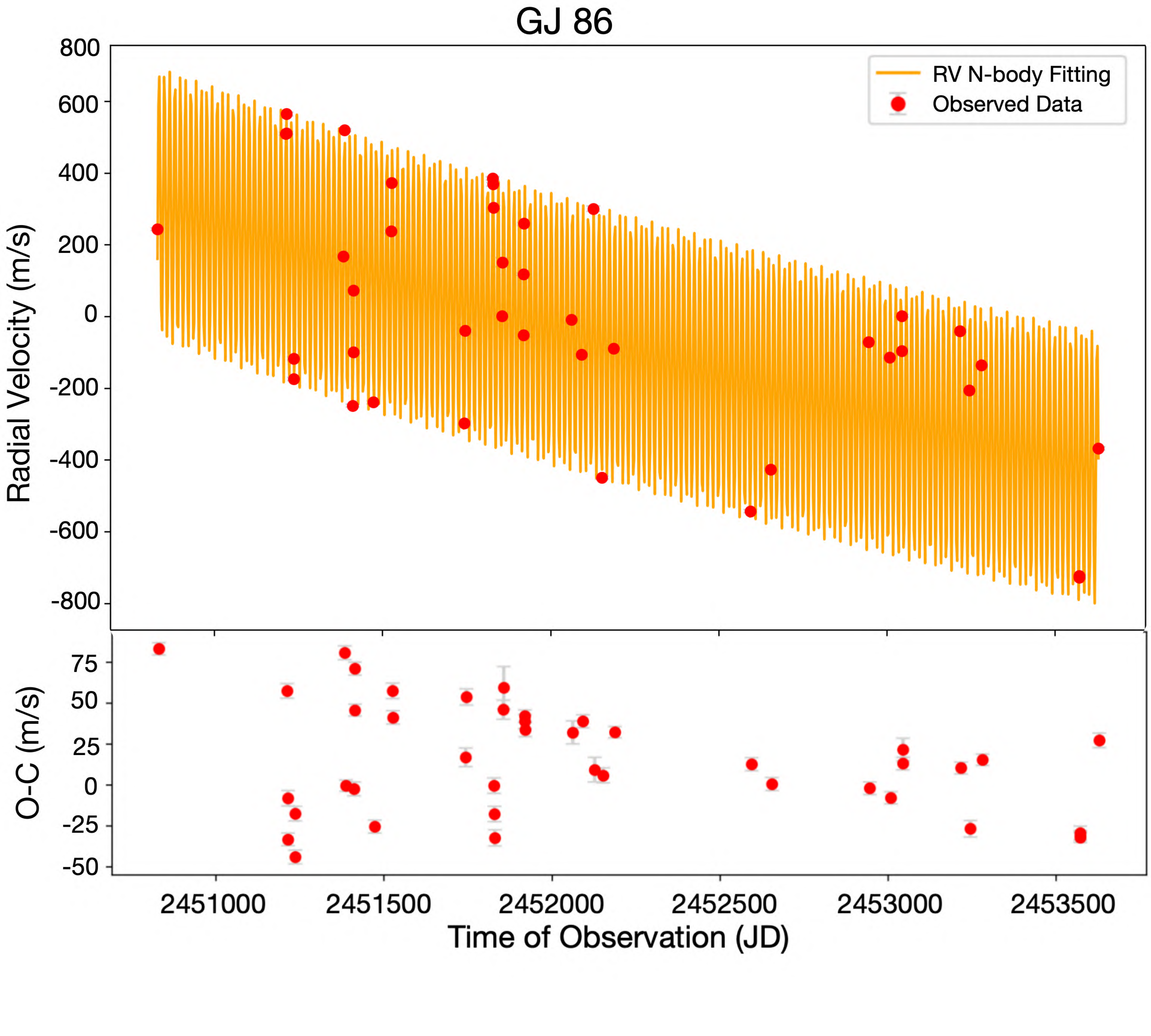}}
         \subfigure[]{\includegraphics[width=0.9\columnwidth,height=7cm]{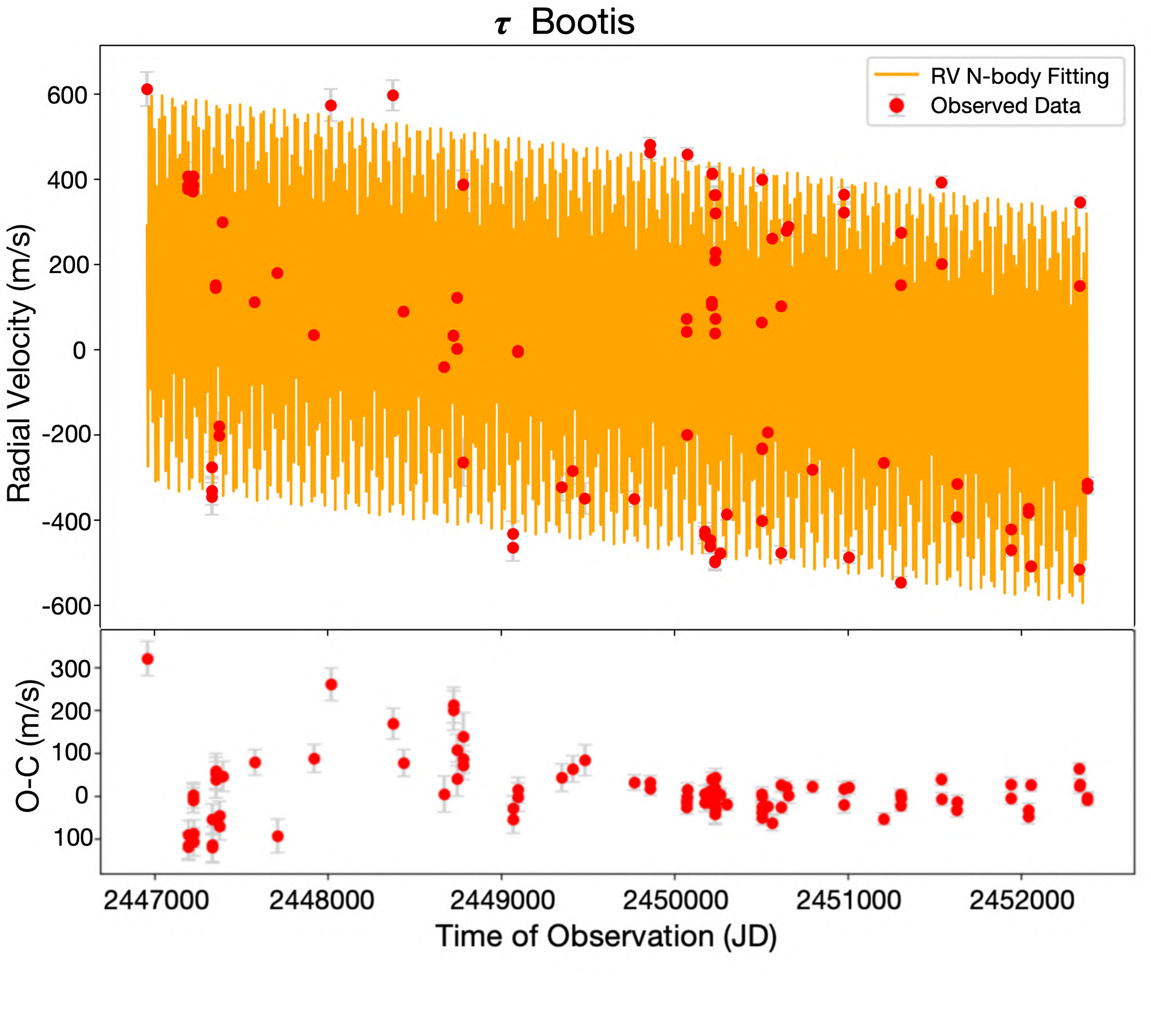}}
  \caption{RV N-body fitting results of GJ 86 Ab B and $\tau$ Bootis Ab B. \textit{Upper panel}: the red dots show published observations of GJ 86 \citep{Fischer2014,Zeng2022} and $\tau$ Bootis \citep{Butler1997,Collier2004}, where the orange solid line denotes the RV curve of the N-body model, the long-term non-periodic trend is shown for both GJ 86 and $\tau$ Bootis with mutual perturbation between the companion and planet b. \textit{Lower panel}: O-C for the N-body model. According to fitting results, GJ 86 Ab is a warm-Jupiter with an orbital period of 15 days, while $\tau$ Bootis Ab is a hot-Jupiter with an  orbital period of 3.31 days.}
  \label{fig:GJ86}
  \end{center}
\end{figure*}

The RV N-body fitting program was developed by combining the N-body integration procedure with the Markov Chain Monte Carlo (MCMC) fitting algorithm. The N-body integrator IAS15 \citep{Rein2015} from REBOUND \citep{Rein2012} was incorporated into the MCMC fitting model to compute the inner perturbed orbit and the theoretical RV signal induced by the planet. Since the orbital elements and the semi-amplitude of the RV ($K$) evolve with time, we select five initial values (at the first observational epoch) for the fitting parameters: $K_1$, $P_1$, $e_1$, $\omega_1$, and $t_{\mathrm{p,1}}$, along with the RV offset ($rv_{\mathrm{offset}}$) as the six fitting parameters. The initial values of $K_{1,0}$, $P_{1,0}$, $e_{1,0}$, $\omega_{1,0}$, and $t_{\mathrm{p,1,0}}$ also serve as the initial conditions for the N-body integration. The MCMC sampler \textit{emcee} \citep{Foreman-Mackey2013} was employed to generate fitting samples and to derive the posterior distribution of the fitting parameters.

The physical parameters of the primary star are summarized in Table \ref{tab:close_binary}. Prior to performing the N-body RV fitting, we applied the Lomb-Scargle (LS) periodogram \citep{Lomb1976,Scargle1982} to identify periodic sinusoidal signals in the RV data. Two representative periodograms are shown in Figure \ref{fig:ls}. Since the observation durations for these systems are much shorter than the orbital period of the stellar companion, the Lomb-Scargle periodogram is primarily sensitive to periodic signals induced by the planet.

\subsection{GJ 86}\label{subsec:3.1}

GJ 86 is a nearby S-type system located 10.9 pc from the Solar System. It hosts the giant planet GJ 86 Ab \citep{Queloz2000}, orbiting at a distance of 0.11 au, while the outer companion is a white dwarf with an eccentric orbit \citep{Mugrauer2005, Lagrange2006}. The orbital elements and masses of both companions were measured by \citet{Zeng2022} using RV data, high-angular-resolution imaging, and absolute astrometry from Hipparcos \citep{Perryman1997} and Gaia \citep{Gaia2016, Brandt2018}. According to \citet{Zeng2022}, the white dwarf companion GJ 86 B has a mass of $m_2 = 0.5425 \pm 0.0042$ $M_{\odot}$, a semi-major axis of $a_2 = 23.7 \pm 0.3$ au, and an eccentricity of $e_2 = 0.429 \pm 0.017$. The orbital inclination of GJ 86 B is derived as ${126.44^{\circ}}^{+0.47}_{-0.49}$, with an ascending node of $234.2^{\circ} \pm 1.0$.

We employed the RV data from the UCLES-Chelle spectrograph \citep{Diego1990} on the Anglo-Australian Telescope (Table 2 of \citet{Zeng2022}).  Here we perform N-body RV fitting to determine the dynamical orbital elements and uncertainty of the minimum planetary mass of GJ 86 Ab, which will be used for the calculation of the true planetary mass in Section \ref{sec:4}. The best-fitting RV curve is shown in Figure \ref{fig:GJ86}(a). The minimum planetary mass derived from the N-body model is 0.2 $M_{\mathrm{Jup}}$ smaller than that from the Keplerian model, which means $m_1$sin$i_1$ was probably overestimated for GJ 86 Ab in the previous work.

\subsection{$\tau$ Bootis}\label{subsec:3.2}

$\tau$ Bootis A is an F6V-type star located at a distance of 15.65 pc, with a stellar mass of $1.320 \pm 0.214 ~ M_{\odot}$. The giant planet orbiting $\tau$ Bootis is one of the best-known exoplanets around nearby stars and was among the first exoplanets to be discovered. $\tau$ Bootis Ab was first discovered by \citet{Butler1997} with an orbital period of just a few days.  \citet{Collier1999} reported measurements of its orbital inclination, but these results were later contested by \citet{Collier2004}. Subsequent studies \citep{Charbonneau1999,Wiedemann2001,Leigh2003,Rodler2010} have attempted to reveal the orbital motion of $\tau$ Bootis Ab through reflected starlight, though its orbital inclination remains uncertain.  Furthermore,  atmospheric thermal emission measurements of $\tau$ Bootis Ab directly measure the planetary radial velocities,  which combined with the stellar RV measurements could also decouple sin($i$) and give the true planetary mass \citep{Brogi2012,Lockwood2014,Pelletier2021,Webb2022,Panwar2024}. \citet{Lockwood2014} derived a planetary orbital inclination of $i$= ${45^{\circ}}^{+3}_{-4}$ and a mass of $m_\mathrm{p}$ = $5.90^{+0.35}_{-0.20}~M_{\mathrm{Jup}}$.

The RV data of $\tau$ Bootis A  is obtained from the Lick Planet Search program \citep{Fischer2014} in the duration of 1987 to 2011\citep{Justesen2019}.  Here,  we conduct N-body RV fitting to obtain  the dynamical orbital elements and the minimum planetary mass of $\tau$ Bootis Ab. The minimum planetary mass derived by the N-body model is less than the Kepler model by  0.49 $M_{\mathrm{Jup}}$ .  The RV curve of the best-fitting results are plotted in Figure \ref{fig:GJ86}(b),  the fitting $\chi^{2} = 4.0914$.

\begin{figure*}
\begin{center}
        \subfigure[]{\includegraphics[width=0.9\columnwidth,height=7cm]{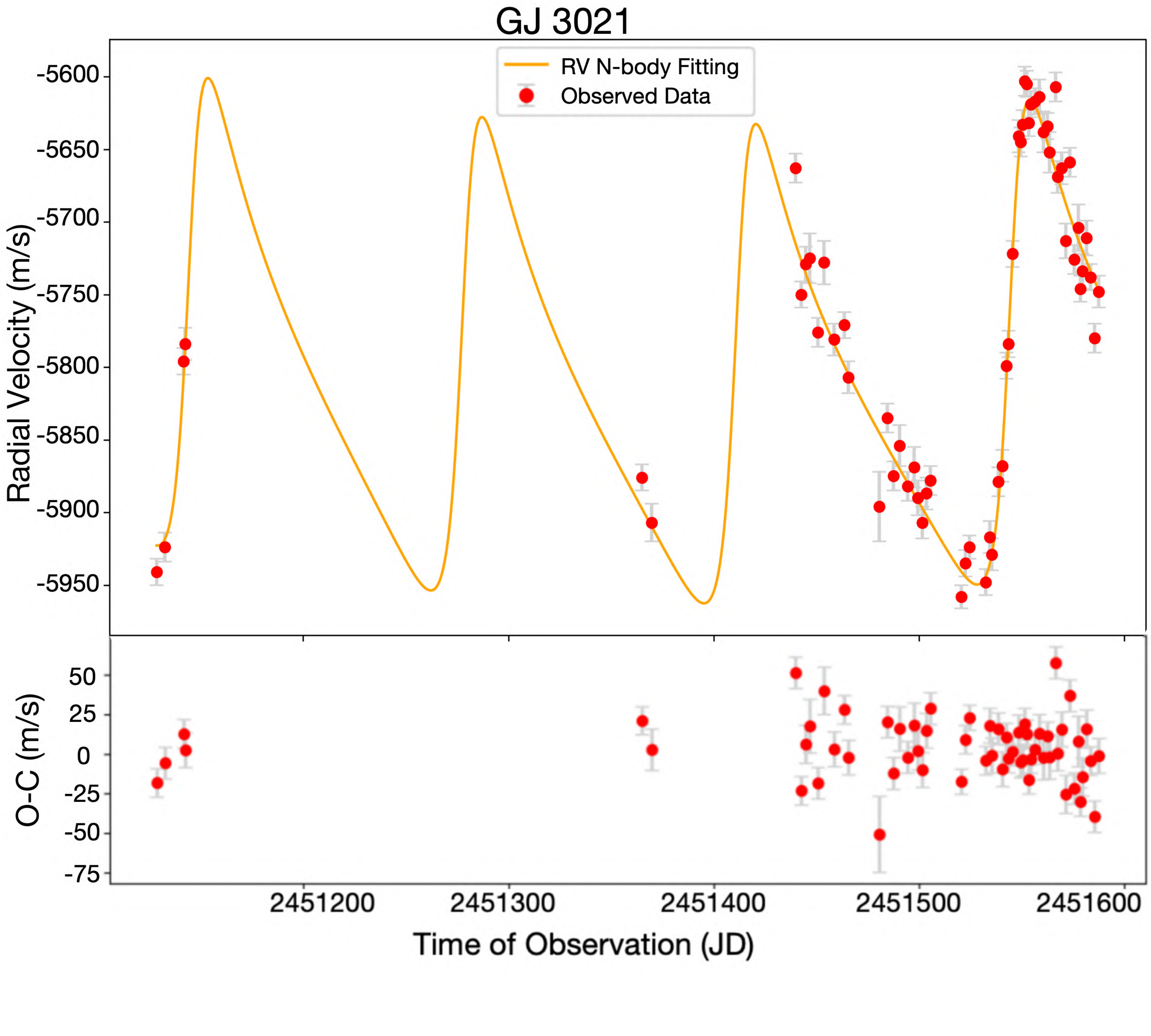}}
         \subfigure[]{\includegraphics[width=0.9\columnwidth,height=7cm]{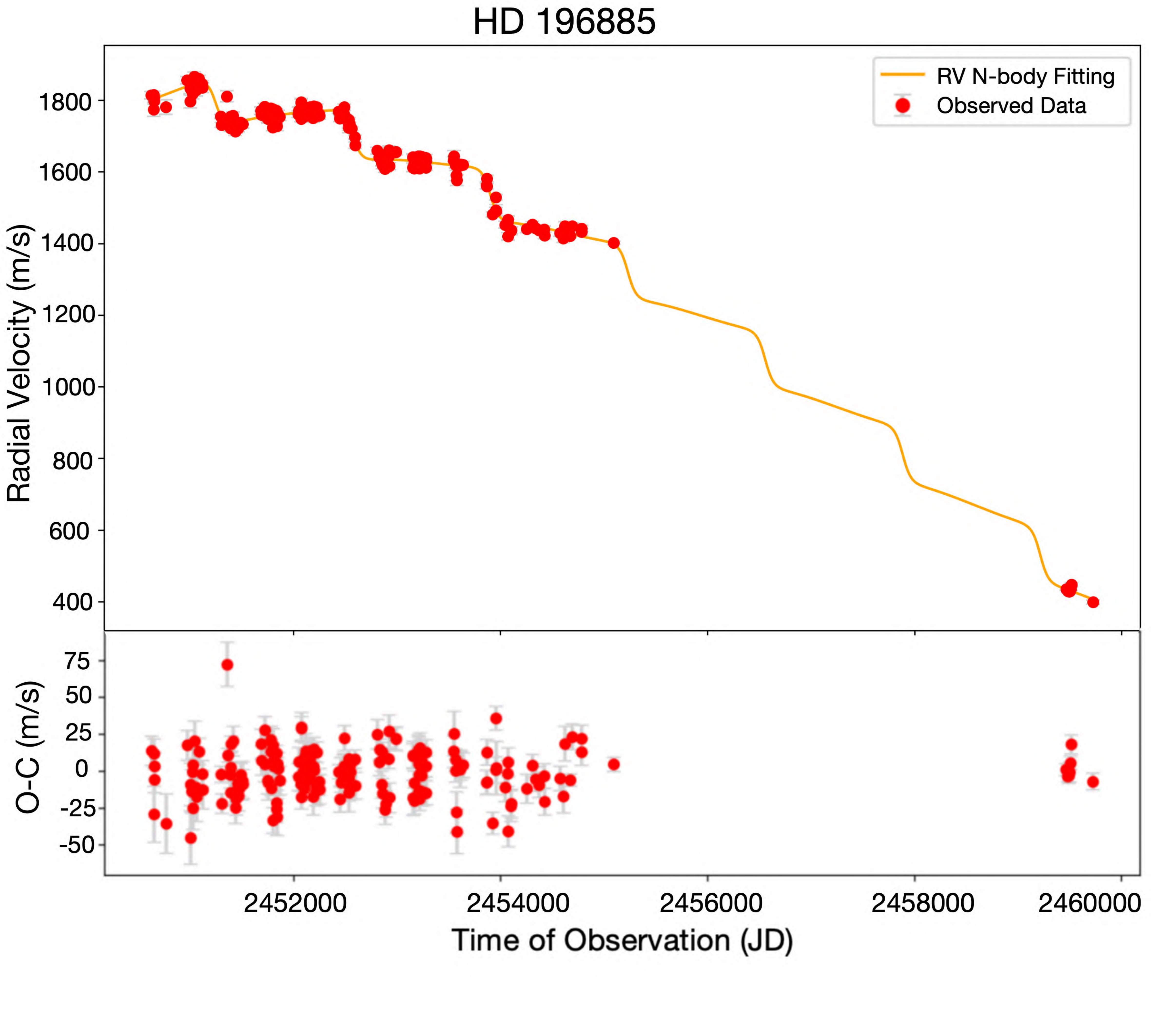}}
  \caption{RV N-body fitting results of GJ 3021 Ab B and HD 196885 Ab B. \textit{Upper panel}: the red dots show published observations of GJ 3021 \citep{Naef2001} and 196885 \citep{Correia2008}, where the orange solid line denotes the RV curve of the N-body model, with mutual perturbation between the companion and planet b. \textit{Lower panel}: O-C for the N-body model, where the residuals of HD 196885 Ab show upward trend of aperiodic signals. According to the fitting results in Table \ref{tab:close_binary_fitting}, GJ 3021 Ab is a super-Jupiter with a 133.4-day orbital period, HD 196885 Ab is a super-Jupiter with a 4-year orbital period.}
  \label{fig:GJ3021}
  \end{center}
\end{figure*}

\begin{figure*}
\begin{center}
        \subfigure[]{\includegraphics[width=0.9\columnwidth,height=7cm]{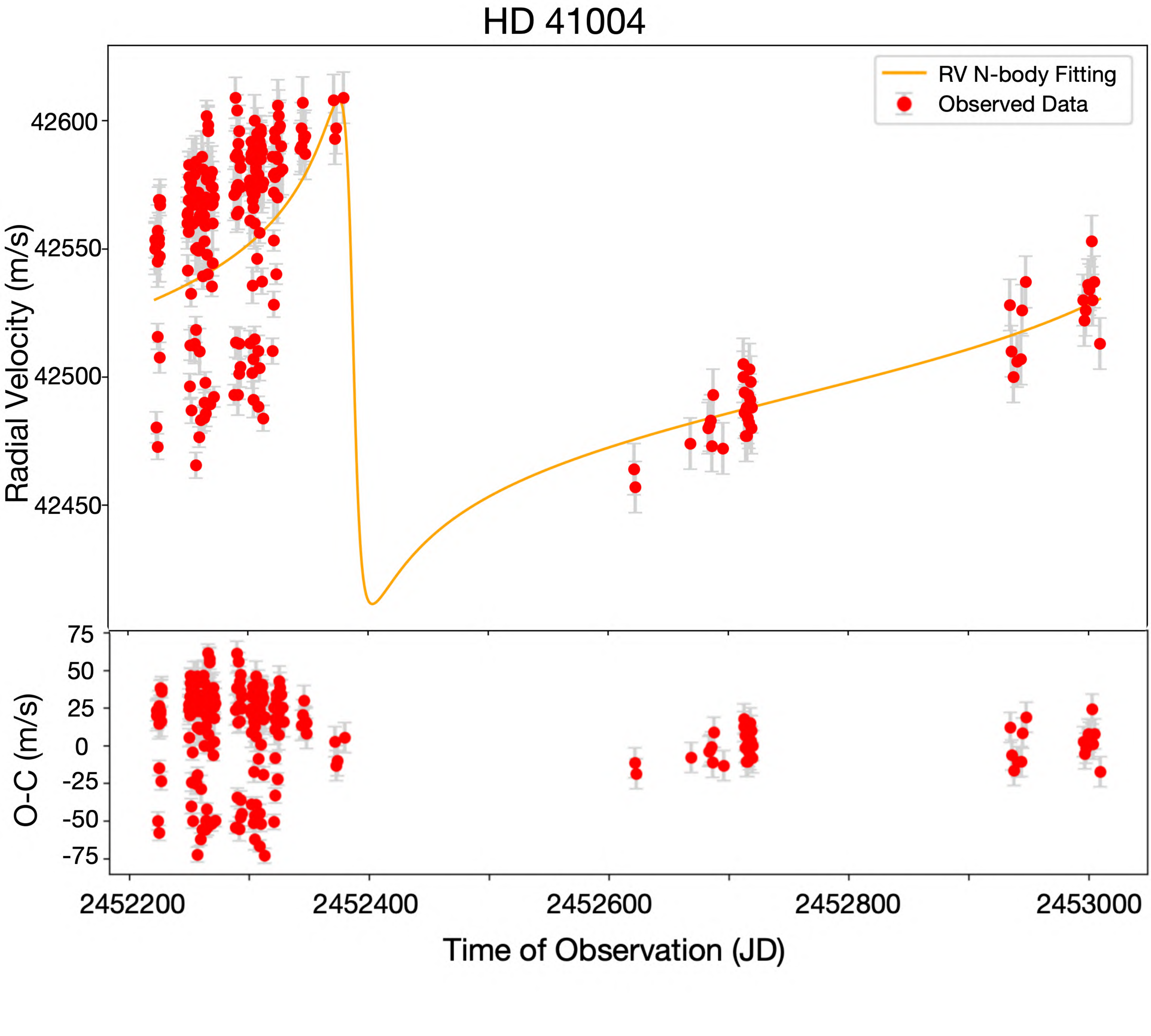}}
         \subfigure[]{\includegraphics[width=0.9\columnwidth,height=7cm]{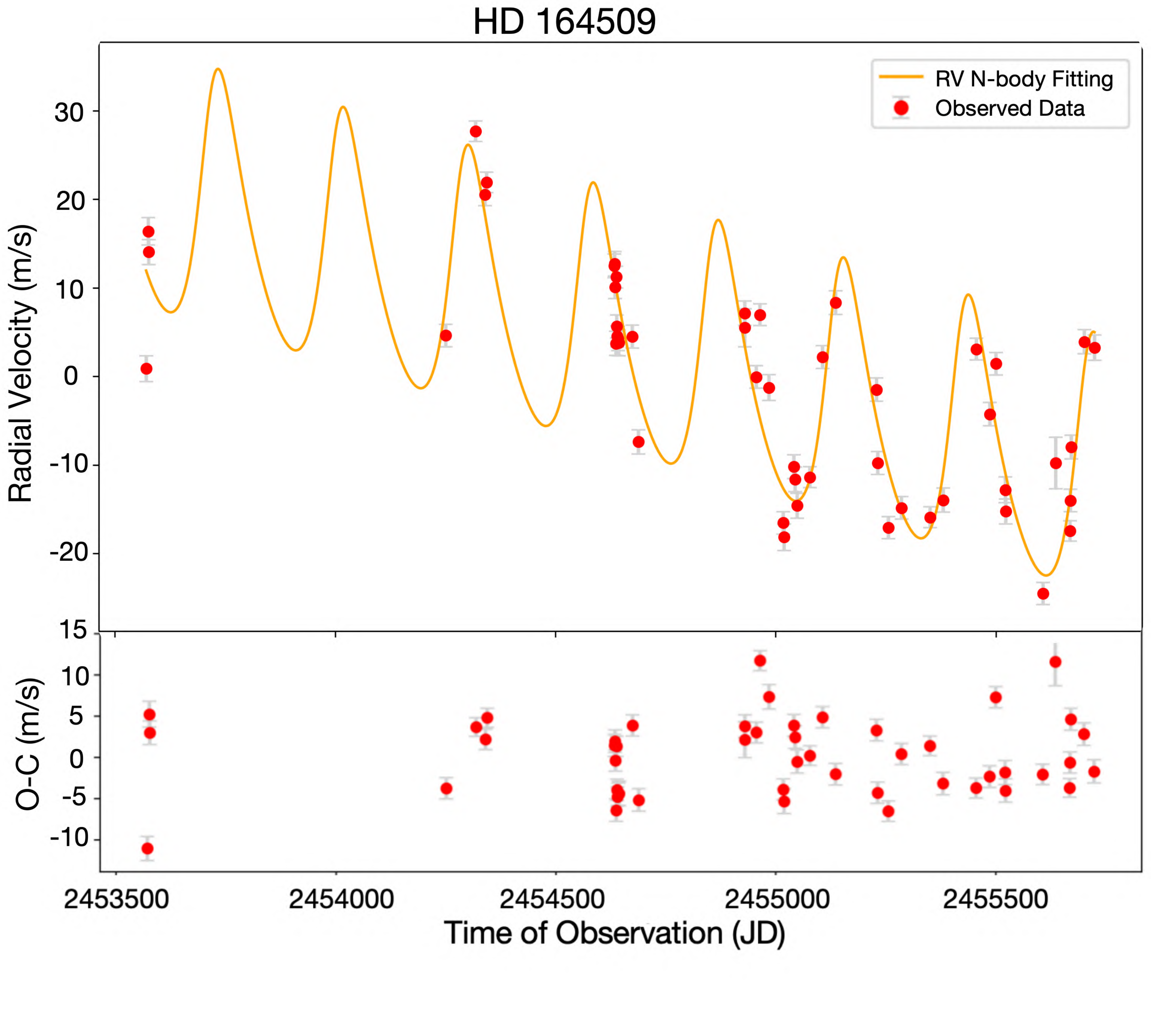}}
  \caption{RV N-body fitting of HD 41004 Ab B and HD 164509 Ab B. \emph{Upper panel}: the red dots show published observations of HD 41004 \citep{Santos2002,Zucker2004} and HD 164509 \citep{Giguere2012}, where the orange solid line denotes the RV curve of the N-body model, with mutual perturbation between the companion and planet b. \emph{Lower panel}: O-C for the N-body model. The results suggest that HD 41004 Ab is a super-Jupiter with a 2.2-year orbital period, while HD 164509 Ab is a super-Neptune with a 283-day orbital period.}
  \label{fig:HD164509}
  \end{center}
\end{figure*}

\begin{table*}
\caption{RV N-body fitting results of S-type planets and the stellar companion in six close binaries}
\label{tab:close_binary_fitting}
\resizebox{2.2\columnwidth}{!}{
\renewcommand{\arraystretch}{1.2}
\centering
\begin{threeparttable}[b]
\begin{tabular} {@{}ccccccc@{}}
\toprule
\multirow{2}*{Parameters} & \multicolumn{2}{c}{GJ 86 Ab B$^{1}$}& \multicolumn{2}{c}{$\tau$ Bootis Ab B$^{2}$} & \multicolumn{2}{c}{GJ 3021 Ab B$^{3}$}\\
\cline{2-3}  \cline{4-5} \cline{6-7}
 & Kepler Model & N-body Model & Kepler Model& N-body Model&Kepler Model& N-body Model \\
 \hline
$K_1(\mathrm{~m} \mathrm{~s}^{-1})$ &$372.81 ^{+49.89}_{-44.75}$& $379.74_{-44.75}^{+49.89}$&$468.42\pm{2.09}$&$462.33^{+3.59}_{-2.69}$&$167\pm{4}$ & $165\pm{2}$ \\
$P_1 $ (days)& $15.78\pm{0.04} $ & $15.53_{-0.02}^{+0.17}$&$3.31\pm({3.3\times 10^{-6}})$&$3.3\pm{9\times 10^{-6}}$&$133.71\pm{0.20}$&$133.40_{-0.01}^{+0.01}$\\
$e_1$ & $0.046 \pm{0.004} $& $0.046_{-0.020}^{+0.010}$&$0.08_{-0.03}^{+0.04}$&$0.02\pm{0.01}$&$0.51\pm{0.02}$&$0.53_{-0.01}^{+0.01}$\\
$\omega_1$ (deg)&$ 270 \pm{4}$&$268_{-17}^{+31}$ &$141.6\pm{25.0}$&$171.4^{+5.7}_{-6.9}$&$290.7\pm{3.0}$&$286.7_{-0.03}^{+0.03}$\\
$T_{\mathrm{p,1}}$ (JD)& $2451147 $&$2451169_{-23}^{+6}$ &$2452658$&$2456402$&$2451546$&$2451531\pm{70}$\\
$m_1 \mathrm{sin}i_1$ ($M_{\text {Jup }}$)& $4.27\pm{0.10}$& $4.09^{+0.56}_{-0.49}$&$4.09\pm{0.14}$& $3.61^{+0.03}_{-0.02}$&$3.37\pm{0.09}$&$3.13\pm{0.06}$\\
\hline
$K_2(\mathrm{~m} \mathrm{~s}^{-1}$) & --&$3127.74^{+170.24}_{-227.29}$& $1446.04_{-463.83}^{+680.92}$& $1297.19^{+125.34}_{-102.15}$&--&$4489.10_{-16.71}^{+8.07}$\\
$P_2$ (days)& $35460_{-680}^{+670}$&$42600_{-7200}^{+9700}$&$883900_{-345900}^{+944900}$&$883800^{+56700}_{-41500}$ &201820&$283850\pm{30}$\\
$e_2$ & $0.43 \pm{0.02}$&$0.31^{+0.19}_{-0.01}$  &$0.87_{-0.03}^{+0.04}$&$0.63\pm{0.02}$&--&$0.87\pm{0.01}$\\
$\omega_2$ (deg)&--&$359.99^{+0.06}_{-1.43}$&$290.7_{-10.0}^{+13.0}$&$104.67^{+2.92}_{-3.52}$&--&$173.77_{-0.04}^{+0.03}$\\
$T_{\mathrm{p,2}}$ (JD)&--&$2446548_{-3227}^{+4281}$& $2461366_{-604}^{+529}$&$2458240^{+4}_{-3}$&--&$2451209_{-62}^{+39}$\\
$m_2 \mathrm{sin}i_2$ ($M_{\odot}$)& $0.5425\pm{0.0042}$&$0.6468^{+0.1254}_{-0.1726}$& $0.49\pm{0.02} $&$0.73^{+0.14}_{-0.11} $&0.13&$0.64\pm{0.01}$\\
\toprule
\multirow{2}*{Parameters}& \multicolumn{2}{c}{HD 196885 Ab B$^{4}$}& \multicolumn{2}{c}{HD 41004 Ab B$^{5}$}& \multicolumn{2}{c}{HD 164509 Ab B$^{6}$}\\
\cline{2-3}  \cline{4-5} \cline{6-7}
&Kepler Model& N-body Model& Kepler Model & N-body Model & Kepler Model& N-body Model \\
 \hline
$K_1(\mathrm{~m} \mathrm{~s}^{-1}$) & $54.40_{-1.50}^{+1.57}$&$54.84_{-1.75}^{+1.87}$& $99\pm{60}$&$108\pm{24}$&$14.2\pm{2.7}$&$14.5_{-0.6}^{+0.8}$   \\
$P_1 (days)$&  $1333.2\pm{3.7}$&$1437.9_{-1.7}^{+1.9}$&$963\pm{38}$&$821\pm{27}$&$282.4\pm{3.8}$&$282.6_{-2.2}^{+0.7}$     \\
$e_1$& $0.48\pm{0.02}$&$0.61\pm{0.02}$&$0.74\pm{0.20}$ &$0.85_{-0.06}^{+0.04}$ &$0.26\pm{0.14}$&$0.26_{-0.03}^{+0.04}$  \\
$\omega_1$ (deg) & $93.2\pm{3.0}$&$110.01\pm{2.29}$&$97\pm{31}$&$75.06^{+14.32}_{-13.75}$ &$324\pm{110}$&$320.28_{-8.02}^{+5.73}$     \\
$T_{\mathrm{p,1}}$ (JD)& $2452571\pm{7}$&$2451170\pm{5}$&$2452425\pm{37}$&$2459792_{-245}^{+244}$&$2455703\pm{30}$&$2455726_{-730}^{+172}$     \\
$m_1 \mathrm{sin} i_1$ ($M_{\text {Jup }}$)&$ 2.98\pm{0.05}$&$2.78_{-0.14}^{+0.15}$&$2.54\pm{0.74}$&$1.98_{-0.64}^{+0.86}$&$0.48\pm{0.09}$&$0.47_{-0.02}^{+0.03}$     \\
\hline
$K_2(\mathrm{~m} \mathrm{~s}^{-1}$) & $2855.12_{-18.48}^{+26.56}$&$2625.55_{-254.68}^{+298.07}$&--&$668.51_{-199.27}^{+219.86}$&--&$1377.36_{-344.96}^{+395.64}$             \\
$P_2$ (days) &$26300\pm{1700}$&$27600_{-2500}^{+3400}$&$38100\pm{10400}$&$50400_{-2300}^{+1800}$ &$66000\pm{5200}$&$131500_{-28900}^{+19300}$                 \\
$e_2$ &$0.42\pm{0.03}$&$0.33_{-0.09}^{+0.08}$&--&$0.82_{-0.12}^{+0.06}$&--&$0.80_{-0.12}^{+0.06}$                                                  \\
$\omega_2$ (deg)&$241.90\pm{3.10}$&$239.50_{-6.30}^{+10.31}$&--&$30.37^{+26.93}_{-20.05}$&--&$2.41_{-2.01}^{+5.16}$                          \\
$T_{\mathrm{p,2}}$ (JD)&$2446362\pm{143}$&$2444472_{-791}^{+735}$&--&$2424945_{-1418}^{+2119}$&--&$2432707_{-15640}^{+5130}$             \\
$m_2 \mathrm{sin}i_2$ ($M_{\odot}$) &$0.45\pm{0.01}$&$0.42_{-0.07}^{+0.08}$&0.4& 0.1&$0.42\pm{0.03}$&$0.21_{-0.09}^{+0.14}$        \\
\hline
\end{tabular}
 \begin{tablenotes}
        \footnotesize
        \item Kepler Model References: 1, \citet{Queloz2000,Zeng2022}; 2, \citet{Wang2011,Justesen2019}; 3, \citet{Naef2001} ; 4, \citet{Chauvin2011,Chauvin2023}; 5,  \citet{Santos2002,Zucker2004}; 6,\citet{Giguere2012}; N-body Model References: this work.
  \end{tablenotes}
  \end{threeparttable}
  }
\end{table*}

\subsection{GJ 3021}\label{subsec:3.3}

GJ 3021 (HD 1237, HIP 1292)\citep{Rocha-Pinto1998,Naef2001,Chauvin2006} is a bright G6-type dwarf star at a distance of 17.62 pc.  We fit the orbit of this system with CORALIE high-resolution RV measurements.  Here we give the dynamical fitting results of GJ 3021's observations.  Since the binary period is much longer than the observation time,  we only give an optimal fit to the orbit of planet b \citep{Naef2001}.  The results are summarized in Table \ref{tab:close_binary_fitting}.

Assuming a prograde orbit for planet b, the orbital inclination is set to $45^{\circ}$. Figure \ref{fig:GJ3021}(a) presents the results of the RV fitting for the GJ 3021 binary system, accounting for the perturbation effects. With an assumed orbital inclination of $45^{\circ}$, the best-fitting results yield $m_1 \sin i_1 = 3.13 \pm 0.06$  $M_{\mathrm{Jup}}$, with a reduced chi-squared value of $\chi^2 = 4.2963$. The minimum planetary mass derived from the N-body model is 0.24 $M_{\mathrm{Jup}}$ less than the value obtained using the Keplerian model.

\subsection{HD 196885}\label{subsec:3.4}

In contrast to GJ 3021, the measurements of the close binary system HD 196885 Ab B span a sufficient duration to simultaneously fit the orbits of both the planet and its companion star, HD 196885 B. The HD 196885 system, located 33 pc from the Solar System, has a primary star of spectral type F8V with a mass 1.3 times that of the Sun. \citet{Correia2008} employed a two-Keplerian model to fit RV data from ELODIE, CORALIE, and CORAVEL observations over 14 years, yielding the first set of orbital solutions. The planet HD 196885 Ab has a minimum mass of 2.96 $M_{\mathrm{Jup}}$, with an orbital period of $P_1 = 3.69 \pm 0.03$ years and an eccentricity of $e_1 = 0.462 \pm 0.026$.

\citet{Chauvin2011} combined available RV data with astrometric data from VLT/NACO observations to refine the orbital inclination and the longitude of the ascending node for star B, yielding $i_{\mathrm{2}} = 116.8^\circ \pm{0.7}$ and $\Omega_{\mathrm{2}} = 79.8^\circ \pm{0.1}$. Planet b, in contrast, is found to orbit in a transiting configuration with an orbital inclination of $89^\circ$. Given that the mutual orbital inclination between the planet and its companion satisfies the Kozai-Lidov excitation condition, the binary stars perturbation will influence the long-term orbital evolution of the system. Using the fitted orbital inclinations for both the planet and companion star, we apply the N-body model to fit the RV data and derive the following orbital elements:
$P_1 = 1437.9^{+1.9}_{-1.7}$ days, $P_2 = 27600^{+3400}_{-2500}$ days,
$e_1 = 0.61 \pm 0.02$, $e_2 = 0.33^{+0.08}_{-0.09}$,
$\omega_1 = {110.0^{\circ}}^{+2.3}_{-2.3}$, $\omega_2 = {239.5^{\circ}}^{+10.3}_{-6.3}$, with $\chi^2 = 2.9063$. The fitted results are listed in Table \ref{tab:close_binary_fitting} and shown in Figure \ref{fig:GJ3021}(b). The minimum planetary mass derived from the N-body model is 0.2 Jupiter masses smaller than the value obtained from the Keplerian model.

\subsection{HD 41004}\label{subsec:3.5}
HD 41004 is G5 main-sequence star at a distance of $52 \pm 3$ pc calculated from the Hipparcos parallax measurement \citep{vanLeeuwen2007}.  The properties of the primary star are listed in Table \ref{tab:close_binary}.  Spectroscopic analysis of HD 41004 yields $T_{\mathrm{eff}} $ = $5922 \pm 44$ K,  [Fe/H] = $0.21 \pm 0.03$ ,  $v$sin$i=  2.4 \pm 0.5$ km/s,  and log$g = 4.44 \pm 0.06$.  \citet{Wittrock2016} derived an absolute visual magnitude of $M_\mathrm{V}$ = 4.64,  and a mass of $1.13 \pm 0.02$ $M_{\odot}$,  and an age of $1.1 \pm 1$ Gyr.

The planet around the star HD 41004 A was discovered by RV measurements \citep{Santos2002,Zucker2004},  the brown dwarf around HD 41004 B was detected by \citet{Zucker2003}.  The semi-major axis of HD 41004 Bb is 0.0177 au,  and its minimum mass is $18.4\pm0.224$  $M_{\mathrm{Jup}}$,  the motion of star A around the barycentric induced by star B and the brown dwarf HD 41004 Bb could be treated as the effect of one star.

\citet{Zucker2004} have obtained a total of 233 precise RV measurements of this system using the CORALIE spectrograph from 2002 to 2004.  Considering the perturbation from the companion star,  we utilize the N-body model to fit the RV data and obtain the orbital elements: $P_1 = 821\pm{27}$ day,  $P_2 = 50400^{+1800}_{-2300}$ day,  $e_1 = 0.85^{+0.04}_{-0.06}$,  $e_2 = 0.82^{+0.06}_{-0.12}$,  $\omega_1 = {75.06^{\circ}}^{+14.32}_{-13.75}$,  $\omega_ 2 = {30.37^{\circ}}^{+26.93}_{-20.05}$, $\chi^{2} = 24.92$.  Here $\chi^{2} $ of HD 41004 Ab Bb is relative greater than other binary system both in the Kepler and N-body model,  since we treat star B and the brown dwarf HD 41004 Bb as one object,  while HD 41004 Bb induces a periodic signal of 1.328 day \citep{Zucker2003}.  We did not include the additional fourth object HD 41004 Bb in our RV model.  The minimum planetary mass derived by the N-body model is 0.6 Jupiter mass less than the Kepler model.  Other results of the fitted RV data are shown in Table \ref{tab:close_binary_fitting},  the RV curve is presented in Figure \ref{fig:HD164509}(a).

\subsection{HD 164509}\label{subsec:3.6}

HD 164509 (HIP 88268) is a G5 main-sequence star located at a distance of $ 52  \pm  3$ pc, as determined from the Hipparcos parallax measurement \citep{vanLeeuwen2007}. The properties of the binary system are listed in Table \ref{tab:close_binary}. Observations of the star began in July 2005 at Keck Observatory using the HIRES spectrometer. A total of 41 observations span a period of five years, with a median velocity error of 1.32 m/s. Spectroscopic analysis of HD 164509 produces effective temperature $T_{\mathrm{eff}} = 5922 \pm 44$ K, metallicity [Fe/H] = $0.21 \pm 0.03$, projected rotational velocity $v \sin i$ = 2.4 $\pm$ 0.5 km/s, and surface gravity $\log g$ = 4.44 $\pm$ 0.06. An absolute visual magnitude of \( M_{\mathrm{V}} = 4.64 \), a mass of \( 1.13 \pm 0.02~M_{\odot} \), and an age of \( 1.1 \pm 1 \) Gyr were given by \citet{Wittrock2016}.

The planet orbiting HD 164509 was discovered in 2011 through RV measurements \citep{Giguere2012}. Previous orbital solutions under the Keplerian model are listed in Table \ref{tab:close_binary_fitting}. To account for the perturbations from the companion star, we utilize the N-body model to fit the RV data and obtain the following orbital elements: $P_1 = 282.58^{+0.67}_{-2.22}$ days, $P_2 = 131484^{+19314}_{-28879}$ days, $e_1 = 0.26^{+0.04}_{-0.03}$, $e_2 = 0.80^{+0.06}_{-0.12}$, $\omega_1 = {320.28^{\circ}}^{ +5.73}_{-8.02}$, $\omega_2 = {2.41^{\circ}}^{ +5.16}_{-2.01}$, and $\chi^2 = 14.21$.  It worth to be caution that $\chi^2$ is also large and the above solutions for the star B may not be the only best fitting results,  because the orbital period of B is 3 times greater than the observation time. We expect future direct imaging or astrometry data to provide more accurate constraints on the orbit of HD 164509 B. The minimum planetary mass derived from the N-body model is 0.1 Jupiter mass smaller than the Keplerian model. Additional fitting results for the RV data are presented in Table \ref{tab:close_binary_fitting}, and the  RV curve is shown in Figure \ref{fig:HD164509}(b).

The deviation in the minimum mass derived from the Keplerian and N-body models, as shown in Table \ref{tab:close_binary_fitting}, reflects the drift in the RV amplitude due to dynamical perturbations. This drift in the RV amplitude, denoted as $K$, in turn provides insights into the mutual inclination between the inner and outer orbits of the binary system. Such analysis can help infer the orbital configuration of the binary when astrometric signals are unavailable. \citet{Matthew2023} utilized measurements from ESPRESSO and HARPS to report a low density for the circumbinary planet TOI-1883 b, which would enable the James Webb Space Telescope (JWST) to perform high signal-to-noise ratio measurements of the chemical composition of the atmosphere of TOI-1338 b. For transiting planets, this methodology can be used to derive a lower planetary density and establish a lower limit for the planetary mass.

\section{Synergy of RV and High-precision Astrometry}\label{sec:4}

Although astrometry alone can give the full orbital parameters of a planet, the longitude of the ascending node $\Omega$ and the argument of periastron $\omega$ are not uniquely given,  since  astrometry detects the projection of the planet's perturbation of the star in the tangential plane of the celestial sphere.  In contrast,  astrometry combined with RV uniquely determines the orbital parameters of the planet, so combining astrometry and RV can solve the 3D orbital solution of the planet. The observational astrometric signal is expressed as follows:

\begin{equation}
\label{equ:alpha}
\begin{aligned}
\alpha &=\frac{m_{\mathrm{p}}}{M_*+m_{\mathrm{p}}} \frac{a}{d} \\ & \approx 5\left(\frac{m_{\mathrm{p}}}{M_{\text {Jup}}}\right)\left(\frac{M_*}{M_{\odot}}\right)^{-1}\left(\frac{a}{5~ \mathrm{au}}\right)\left(\frac{d}{1~\mathrm{pc}}\right)^{-1} mas  \\
& \approx 3\left(\frac{m_{\mathrm{p}}}{M_{\oplus}}\right)\left(\frac{M_*}{M_{\odot}}\right)^{-1}\left(\frac{a}{1~ \mathrm{au}}\right)\left(\frac{d}{1~\mathrm{pc}}\right)^{-1} \mu as
\end{aligned}
\end{equation}

 \begin{figure*}
\begin{center}
        \subfigure[]{\includegraphics[width=1.0\columnwidth,height=7.5cm]{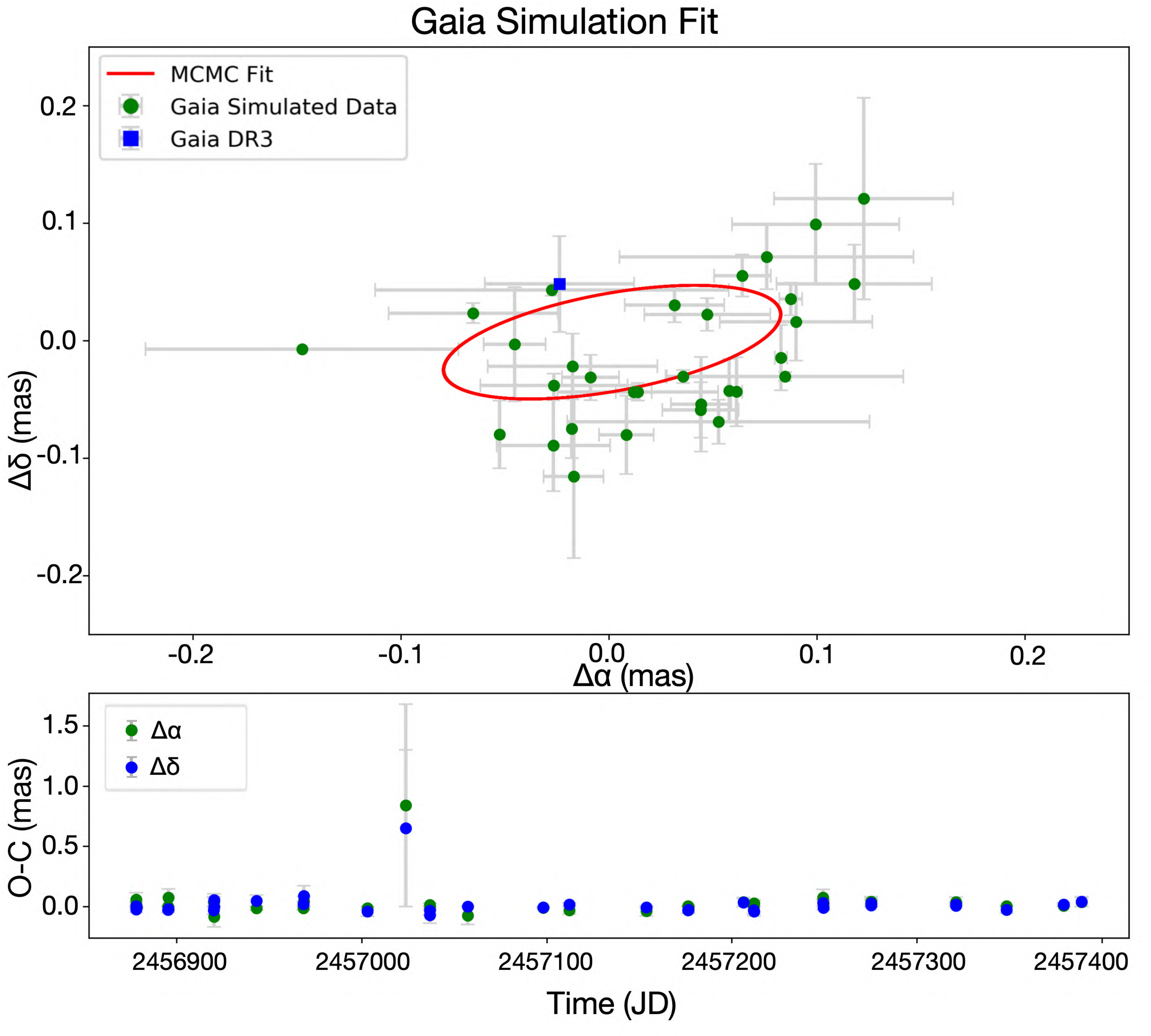}}
         \subfigure[]{\includegraphics[width=1.0\columnwidth,height=7.5cm]{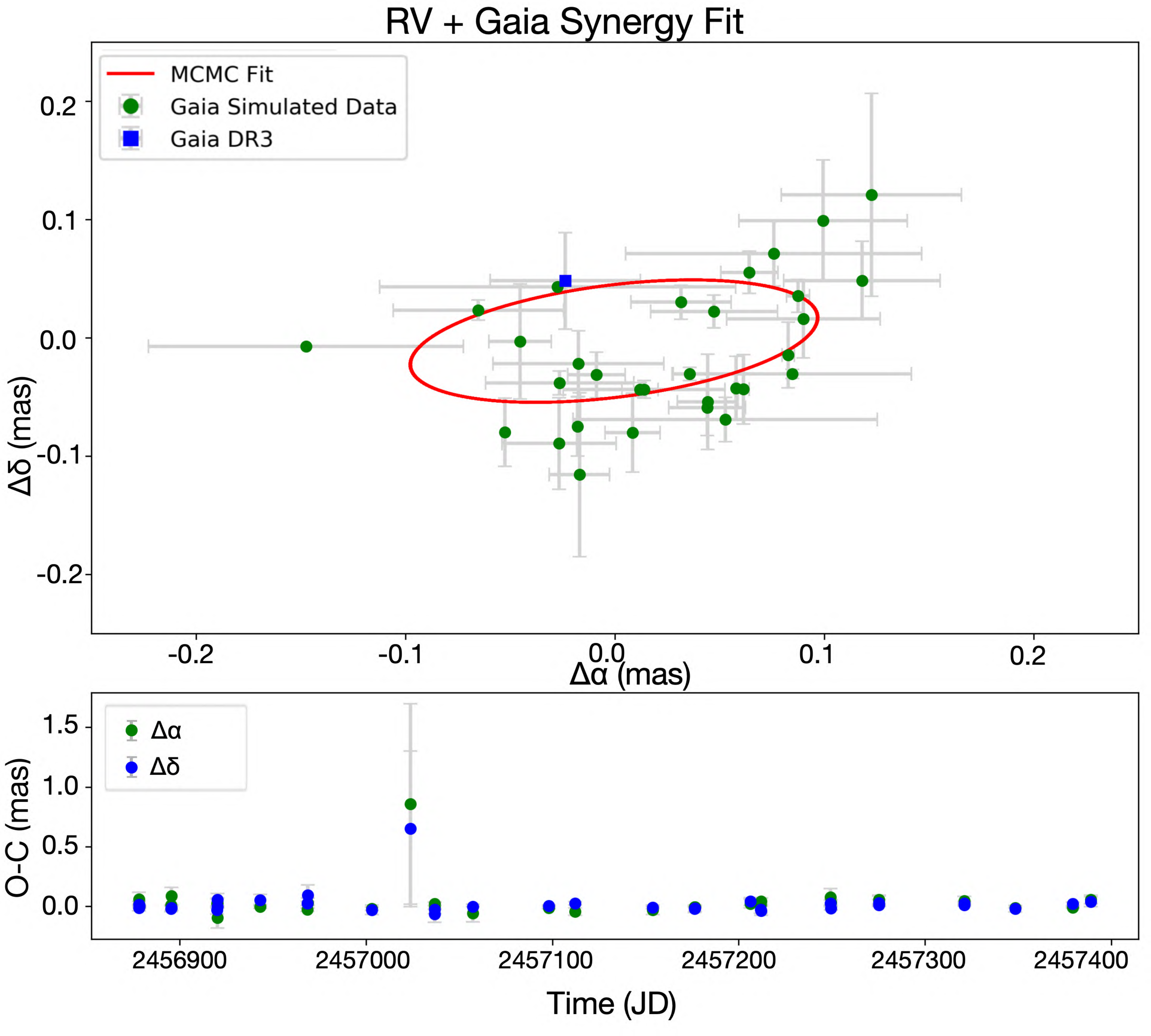}}
  \caption{The MCMC best-fitting results with Gaia simulation data. Green dots are 33 simulated data of GJ 86 generated by Gaia DR3 with accuracy of 57.8 $\mu$as, the blue square is the measurement of star position by Gaia DR3. The red curve is the MCMC fitting result of GJ 86 Ab orbit, which denotes the theoretical variations of star position induced by the planet orbit. \textit{Left panel}: the Gaia Astrometry-only fitting result, the best-fitting  inclination of GJ 86 Ab is $i_1 = 54.32^{\circ}$. \textit{Right panel}: the RV + Gaia Synergy fitting result.  The best-fitting  inclination of GJ 86 Ab is $i_1 = 57.15^{\circ}$. The O-C panel is given below, which equals fitted orbit subtracted with simulated right ascension $\alpha$ and declination $\delta$. Green dots with error bar are $\Delta \alpha$, blue dots are $\Delta \delta$.}
  \label{fig:GJ86_gaia}
  \end{center}
\end{figure*}

\begin{figure*}
\begin{center}
        \subfigure[]{\includegraphics[width=1.0\columnwidth,height=7.5cm]{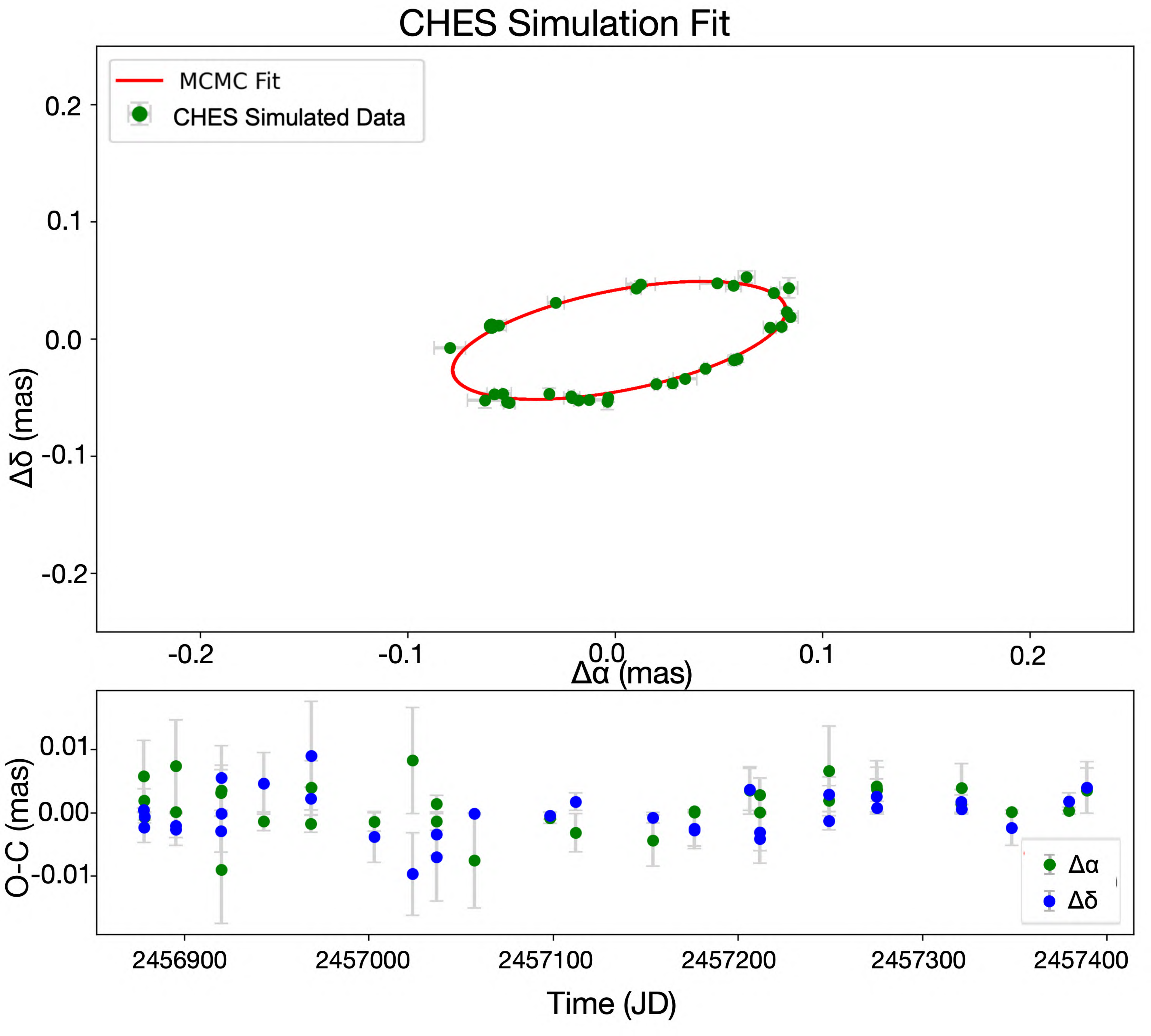}}
         \subfigure[]{\includegraphics[width=1.0\columnwidth,height=7.5cm]{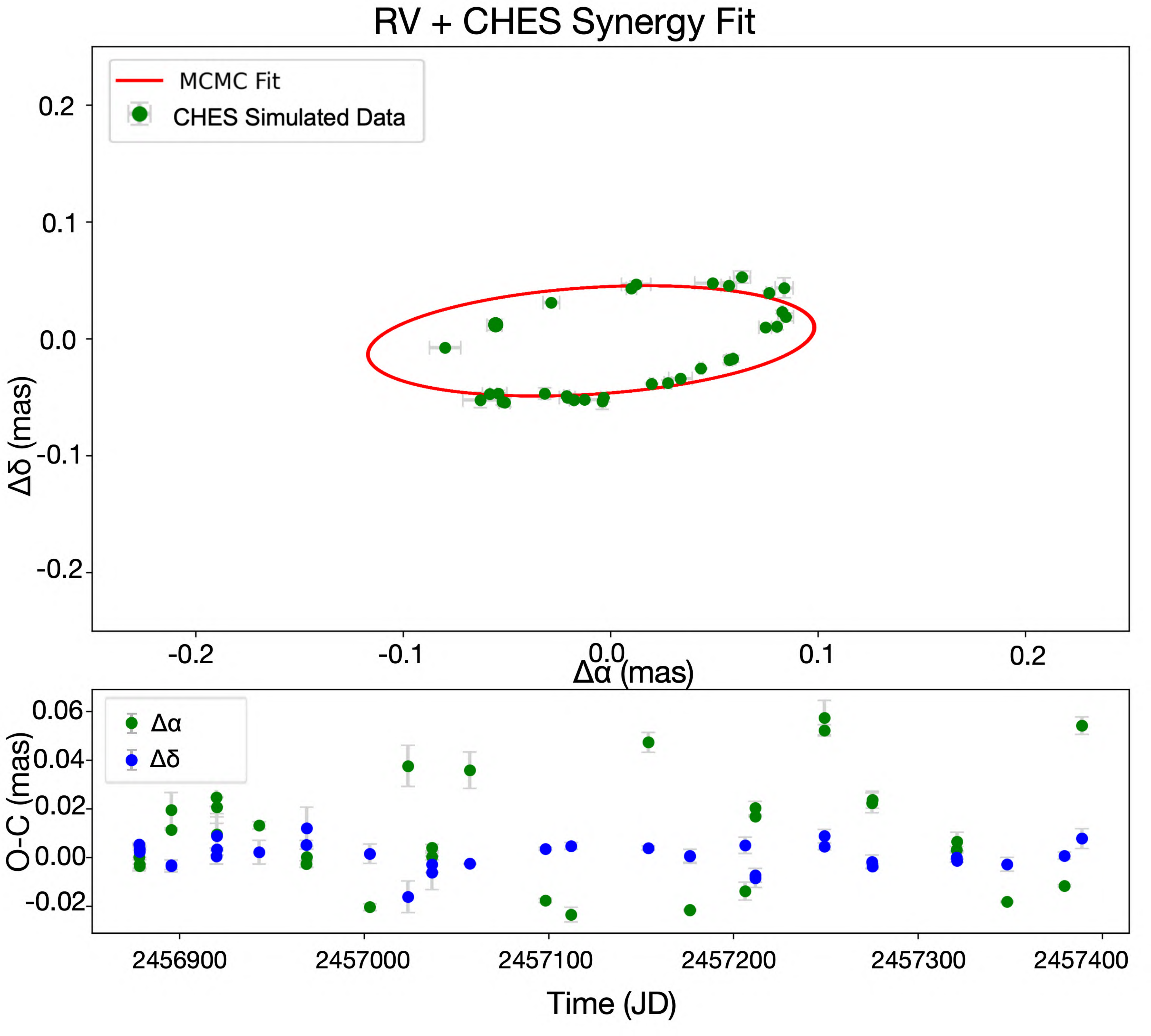}}
           \caption{The MCMC best-fitting results with CHES simulation data. Green dots are 33 simulated data of GJ 86 generated by CHES observation strategy \citep{Ji2024,Tan2024}, with accuracy of 1 $\mu as$, red curves represent the variation of star position induced by the fitted planet orbit.  \textit{Left panel}: the CHES Astrometry-only fitting result. The best-fitting  inclination of GJ 86 Ab is $i_1 = 54.96^{\circ}$. \textit{Right panel}: the RV + CHES Synergy fitting result. The best-fitting  inclination of GJ 86 Ab is $i_1 = 50.00^{\circ}$. The dots in the panel below refer to the same parameters as in Figure~\ref{fig:GJ86_gaia}.}
  \label{fig:GJ86_ches}
  \end{center}
\end{figure*}

The synergy between RV and astrometry requires the retrieval of 11 parameters, namely: $K$, $P$, $e$, $\omega$, $T_{\mathrm{p}}$, $rv_{\mathrm{offset}}$, $m_{\mathrm{p}}$, $i$, $\Omega$, $\mu_{\alpha}$, and $\mu_{\delta}$. To simplify the simulation, we use the following relationship between the astrometric signal  and the RV semi-amplitude \citep{Pourbaix2000}:

\begin{equation}
\label{equ:alpha2}
\begin{aligned}
\frac{\alpha \sin i}{\varpi}=\frac{P K \sqrt{1-e^2}}{2 \pi}
\end{aligned}
\end{equation}

RV observation data and the  astrometry data are merged as one input data list,  the observation error is put in the second column of the list.  We construct the theoretical RV function and the astrometry function in the fitting model of \textit{emcee} program,  then append RV and  astrometry as one data list to compare with the input observation data list.  \textit{emcee} sampler estimates the best fitting model by numerically optimizing the likelihood function.  When calculating the posterior probability distribution of fitting parameters in the RV + Astrometry synergy method,  the likelihood function is expressed as:

\begin{equation}
\label{equ:alpha2}
\begin{aligned}
\ln \mathcal{L}= \ln \mathcal{L}_{RV} + \ln \mathcal{L}_{\mathrm{Astrometry}}
\end{aligned},
\end{equation}
where \( \ln \mathcal{L} \) is the total likelihood function, and \( \ln \mathcal{L}_{\mathrm{RV}} \) and \( \ln \mathcal{L}_{\mathrm{Astrometry}} \) represent the likelihood functions for the RV and astrometric models, respectively.

First, according to the theoretical astrometric signal equations (Equation \ref{equ:alpha}), the time-series evolution of $\alpha$ is primarily governed by the variation in the planetary orbital semi-major axis. Over the 20-year observation baseline, the variations in the semi-major axis are too small to be distinguished from the astrometric simulation data. Particularly, the tiny effect induced by the perturbation is negligible for planets in close binaries, as the N-body model is predominantly influenced by secular perturbations. Additionally, we demonstrate that the parameter space of planetary mass with 99\% detectability remains largely unaffected by the perturbing companions mass or the mutual inclination between the inner and outer planetary orbits.

To determine the true masses of planets in close-binary systems, we combine astrometric data with RV measurements. As is well known, Gaia Data Release 2 (DR2) \citep{Gaia2018} was first made available in 2018, with the subsequent release of Gaia Early Data Release 3 (EDR3) \citep{Gaia2021}. For most sources in the Gaia catalog, time-series astrometric data are not yet available but are expected to be released in Gaia Data Release 4. In Section \ref{subsec:4.1}, we fit the GJ 86 system using available RV data and simulated astrometric data extrapolated from Gaia DR3. In Section \ref{subsec:4.2}, we compare the Gaia simulation fitting results with higher precision astrometric data obtained from the CHES simulation.

\subsection{Synergy with Gaia Simulation data}
\label{subsec:4.1}

Here we utilize the orbital elements of GJ 86 Ab B derived from N-body RV fitting to generate  astrometric simulation data with Gaia precision $\sigma$ = 57.8 $\mu as$.  More than $50\%$ of the S-type exoplanets are detected by radial velocities and have unknown orbital inclinations. In order to get the inclination with the highest confidence, we set other orbital elements as known constants, and varying only the a priori value of the initial orbital inclination,  the orbital inclination of the S-type planets are set in the range of ($5^{\circ}$ -- $85^{\circ}$).  Position of the target is extrapolated from Gaia DR3 at epoch of JD-2457389,  $\alpha =32.6229^{\circ} \pm 0.0359$,  $\delta = -50.8209^{\circ}  \pm 0.0408$, $\mu_{\alpha}$ = $2125.416 \pm{0.048}$ mas/yr,  $\mu_{\delta}$ = $637.975\pm{0.062}$ mas/yr,  the parallax $\pi$ =  $92.9251 \pm 0.0461$ mas . The observation epoch of simulated data are between Gaia DR1 and Gaia DR3, and are inferred from \href{https://gaia.esac.esa.int/gost/}{\textit{Gaia Observation Forecast Tool}}.  The number of observations for the planet is set as 33.

We employ both the Keplerian Astrometry-only model and the RV + Astrometry synergy model to derive the true planetary mass. The fitting results for $i_1 = 55^{\circ}$ from both the Astrometry-only model and the RV + Astrometry synergy model are presented in Table \ref{tab:close_binary_fitting_as}.

The astrometric signal induced by the planetary orbit, simulated with Gaia precision, is presented in Figure \ref{fig:GJ86_gaia}. In the case shown here, the orbital inclination is assumed to be $55^{\circ}$. The Astrometry-only fitting result is shown in Figure \ref{fig:GJ86_gaia} (a), where the fitted proper motion values are $\mu_{\alpha} = 2125.4161$ mas/yr and $\mu_{\delta} = 637.9760$ mas/yr. The fitting error is $1.56 \times 10^{-6}$ mas/yr, and the best-fitting inclination of GJ 86 is $54.32^{\circ}$, with a minimum $\chi^2 = 0.0655$. The RV + Astrometry synergy fitting result is shown in Figure \ref{fig:GJ86_gaia} (b). By comparing the $\chi^2$ values of all synergy fitting cases with $i_1 = 5^{\circ} - 85^{\circ}$, we conclude that the best-fitting orbital inclination of GJ 86Ab is constrained within the range of $40^{\circ} - 60^{\circ}$. As shown in Table \ref{tab:close_binary_fitting_as}, the planetary mass derived from the synergy method is $0.56 ~ M_{\mathrm{Jup}}$ larger than that from the Astrometry-only method.

\subsection{Synergy with CHES Simulation Data}
\label{subsec:4.2}

GJ 86 has been selected as one of the candidates for the CHES mission \citep{Ji2024,Bao2024a,Tan2024}, which aims to observe nearby solar-type stars in order to search for terrestrial planets within habitable zones at ultra-high resolution via astrometry. We utilize the orbital elements derived from N-body RV fitting to generate astrometric simulation data, assuming CHES precision of $\sigma = 1 ~ \mu\text{as}$.

The process from CHES observations to planetary signal data can be described as follows: by measuring the temporal variations in the angular distances between a target star and various reference stars within the field of view (FOV). Models are then constructed to account for the effects of proper motion,  parallax, and planetary perturbations on these changes.  Subsequently,  the components of planetary gravitational effect in each angular direction are extracted from the angular distances. By utilizing the spatial distribution of reference stars in the FOV, the 3-D planetary orbital effects in different angular directions are paired and combined to reconstruct a 2-D projection of the stellar motion induced by the planet onto the observational plane (Tan et al., in prep.).

\begin{table*}
\centering
\caption{Fitting results of the parameters of GJ 86 Ab with Astrometry and the RV+Astrometry model.}
\renewcommand\arraystretch{1.2}
\label{tab:close_binary_fitting_as}
\begin{tabular} {ccccccc}
\toprule
\multirow{3}*{Parameters} & \multicolumn{2}{c}{Gaia Simulation}& \multicolumn{2}{c}{CHES Simulation} \\
\cline{2-3}  \cline{4-5}
 & Astrometry & RV+Astrometry & Astrometry & RV+Astrometry\\
 \hline
 $m_1 $ ($M_{\text {Jup }}$)& $5.16^{+0.15}_{-0.14}$ & $5.73 \pm{0.01}$& $5.26\pm{0.02}$ & $5.90\pm{0.01}$\\
$P_1$ (days)&$15.76\pm{0.01}$&	$15.77 \pm{0.01}$& $15.76\pm{0.01}$&   $15.77\pm{0.01}$	\\
$e_1$ & $0.046^{+0.010}_{-0.013}$& $ 0.056\pm{0.003}$	&$0.048\pm{+0.003}$ & $0.088\pm{0.001}$\\
$\omega_1 $ (deg) & $272.19^{+16.40}_{-18.16}$ &  $ 297.05 \pm{2.41}$& $269.71^{+2.30}_{-2.38}$ &   $359.99\pm{0.01}$\\
$T_{\mathrm{p,1}}$ (JD)& $2452804\pm{1}$& $2452803$&$2452804$ &$2452806$\\
$i_1$ (deg) & $54.32^{+2.01}_{-1.86}$ & $50.01^{+0.02}_{-0.01}$& $54.96^{+0.32}_{-0.30}$&$50.00\pm{0.01}$\\
$\Omega_1$ (deg)& $48.30^{+3.58}_{-3.40}$ & $57.15^{+0.69}_{-0.70}$ & $45.44^{+0.38}_{-0.40}$ &	$73.33^{+0.10}_{-0.11}$\\
\hline
\end{tabular}
\end{table*}

The astrometry fitting results derived from CHES simulation data are presented in Figure \ref{fig:GJ86_ches}. The Astrometry-only fitting result is shown in Figure \ref{fig:GJ86_ches} (a), where the best-fitting inclination of GJ 86 is found to be $54.9630^{\circ}$, with a minimum $\chi^2$ of 1.5300. The RV + Astrometry synergy fitting result is plotted in Figure \ref{fig:GJ86_ches} (b). According to Table \ref{tab:close_binary_fitting_as}, the planetary mass derived from the synergy method exceeds that obtained from the Astrometry-only method by $0.65~M_{\mathrm{Jup}}$, which is in close agreement with the Gaia simulation results. Furthermore, the orbital element fitting results from the Astrometry-only method with CHES simulation data are most consistent with the RV fitting results presented in Section \ref{sec:3}.

As discussed in Section \ref{sec:2}, the RV data resolution is significantly more sensitive to the N-body model, whereas the N-body effects can be weakened in the astrometry fitting process if the semi-major axis evolution timescale. is much longer  than the observation baseline.  Consequently, the goodness-of-fit of the RV + Astrometry synergy model does not necessarily outperform that of the Astrometry-only model. Moreover, the goodness-of-fit in the N-body framework is influenced by the long-term stability of the binary system, which requires the stability timescale to be much longer than the observation duration.

\section{Summary}\label{sec:5}
This work aims to integrate dynamical models with the MCMC orbital fitting method to improve the orbital solutions for the planets in the binary systems. The study is motivated by advancements in next-generation RV spectrometers (ESPRESSO, HARPS) and the forthcoming high-precision astrometry mission CHES. By leveraging the dynamical evolutionary features of MMRs systems and close binaries, we further substantiate the conclusion that previously neglected error signals in the planetary observations may harbor significant dynamical information. The magnitude of these dynamical effects is within a range that can be probed and characterized using high-precision RV and astrometric data.

Section \ref{sec:2} concludes that the detection efficiency of Earth-like planets in the $m_1$sin$i_1$--$a_1$ parameter space increases with the presence of a stellar companion, which results directly from the drift of $ K $.  This drift of $ K $ can be detected by the RV criteria of 1 m/s.  For example,  in Figure \ref{fig:rv_efficiency_binary},  among all cases with $\omega = 0^{\circ}$ to $ 360^{\circ}$,  if the proportion of cases where  $ K $  drifts upward beyond the critical value for 50\% detection efficiency $K_{\mathrm{50}} $  is greater than those where  $ K $  drifts downward, then the overall detection probability increases in the parameter space.  Otherwise,  the detection probability decreases.

In Section \ref{sec:3},  we develop an N-body fitting code to perform RV fitting. The dynamical fitting results in Section \ref{sec:3} yield a smaller planetary mass compared to the Keplerian model results.  The deviation in the minimum mass between the Kepler and N-body models reflects the drift of the RV semi-amplitude due to dynamical interactions. This drift,  in turn, provides insight into the mutual inclination relationship between the inner and outer orbits of the binary system.  Such information helps infer the orbital configuration of the binary when the sky survey signal is not available.

In Section \ref{sec:4},  we respectively combine the Gaia and CHES simulation data with available RV data in selected S-type planetary systems,  to constrain the uncertainty of the planetary mass.  This section presents a novel approach for constraining the range of planetary inclinations when epoch astrometric data is unavailable.   We utilize the RV fitting results in Section \ref{sec:3} to generate Astrometry simulation data,  then we could conduct the RV+Astrometry synergy fitting for all assumed cases of $i_1 = 5^{\circ} - 85^{\circ}$.

By comparing the \( \chi^2 \) values of all synergy fitting cases with \( i_1 = 5^\circ - 85^\circ \), we conclude that the best-fitting orbital inclination for GJ 86 Ab is constrained to the range \( 40^\circ - 60^\circ \) in both the RV+Gaia and RV+CHES synergy analyses. The planetary mass derived from the synergy method is approximately \( 0.6~M_{\mathrm{Jup}} \) greater than that obtained from the astrometry-only method for GJ 86 Ab (see Table \ref{tab:close_binary_fitting_as}).  Furthermore, the fitting results from the astrometry-only method using CHES simulation data are in good agreement with the RV results presented in Section \ref{sec:3}.

Comparing $\chi^2$ of all synergy fitting cases with $i_1 = 5^{\circ} - 85^{\circ}$,  we conclude that the best-fitting orbital inclination of GJ 86 Ab is constrained in the range of $40^{\circ}\sim60^{\circ}$ both in RV + Gaia synergy and RV + CHES synergy.  According to Table \ref{tab:close_binary_fitting_as},  the planetary mass derived from the synergy method is greater than the Astrometry-only method by $\sim  0.6~M_{\mathrm{Jup}}$ for GJ 86 Ab. The fitting results of Astrometry-only method with CHES simulation data is well consistent with the RV fitting results.

Except for conducting the RV+Astrometry synergy fitting,  we calculated the astrometry detection probability using the N-body model independently as well.  We observed no significant drift in the astrometric signal or any notable increase or decrease in the detection probability.  This could be attributed to the fact that the planet-induced astrometric signal of the central star primarily depends on the semi-major axis and planetary mass.  Neither the MMRs nor the secular evolution of the binary star leads to significant changes in the orbital semi-major axis.  Consequently,  the current accuracy of astrometric measurements is not sufficient to detect the weak signals from orbital dynamical evolution.  However,  there is potential for future high-precision astrometry data,  such as from CHES,  to provide similar dynamical insights as the RV N-body model.  Further combination with ground-based high-precision RV spectrograph could improve our understanding of  configuration of planetary systems and better constrain the dynamical mass of planets.

\section*{Acknowledgements}
We thank the anonymous reviewer for the insightful comments and suggestions that improved the quality of the original manuscript. This work is financially supported by the National Natural Science Foundation of China (Grant Nos.12033010, 11773081, 12473076), the Strategic Priority Research Program on Space Science of the Chinese Academy of Sciences (Grant No. XDA 15020800), the Natural Science Foundation of Jiangsu Province (Grant No. BK20221563), the Foreign Expert Project (Grant No. S20240145), and the Foundation of Minor Planets of the Purple Mountain Observatory.

\bibliography{ms}{}
\bibliographystyle{aasjournal}

\end{CJK*}

\end{document}